\documentclass[preprint]{aastex}
\usepackage{mkfig}

\begin{document}

\title{New Observational Constraints on the Winds of M Dwarf Stars\altaffilmark{1}}

\author{Brian E. Wood\altaffilmark{2}, Hans-Reinhard M\"{u}ller\altaffilmark{3},
  Seth Redfield\altaffilmark{4}, Fallon Konow\altaffilmark{4},
  Hunter Vannier\altaffilmark{4}, Jeffrey L. Linsky\altaffilmark{5},
  Allison Youngblood\altaffilmark{6}, Aline A. Vidotto\altaffilmark{7},
  Moira Jardine\altaffilmark{8}, Juli\'{a}n D. Alvarado-G\'{o}mez\altaffilmark{9},
  Jeremy J. Drake\altaffilmark{10}}
\altaffiltext{1}{Based on observations made with the NASA/ESA Hubble
  Space Telescope, obtained at the Space Telescope Science Institute,
  which is operated by the Association of Universities for Research
  in Astronomy, Inc., under NASA contract NAS 5-26555.  These
  observations are associated with program GO-15326.}
\altaffiltext{2}{Naval Research Laboratory, Space Science Division,
  Washington, DC 20375, USA; brian.wood@nrl.navy.mil}
\altaffiltext{3}{Department of Physics and Astronomy, Dartmouth College,
  Hanover, NH 03755, USA}
\altaffiltext{4}{Astronomy Department and Van Vleck Observatory, Wesleyan University,
  Middletown, CT 06459-0123, USA}
\altaffiltext{5}{JILA, University of Colorado and NIST, Boulder, CO 80309-0440, USA}
\altaffiltext{6}{Laboratory for Atmospheric and Space Physics, University of
  Colorado, Boulder, CO 80303, USA}
\altaffiltext{7}{School of Physics, Trinity College Dublin, College Green,
  Dublin-2, Ireland}
\altaffiltext{8}{SUPA, School of Physics and Astronomy, North Haugh, St. Andrews,
  Fife KY16 9SS, UK}
\altaffiltext{9}{Karl Schwarzschild Fellow, Leibniz Institute for Astrophysics
  Potsdam, An der Sternwarte 16, D-14482 Potsdam, Germany}
\altaffiltext{10}{Smithsonian Astrophysical Observatory, 60 Garden St.,
  Cambridge, MA 02138, USA}

\begin{abstract}

     High resolution UV spectra of stellar H~I Lyman-$\alpha$ lines from the
{\em Hubble Space Telescope} (HST) provide observational constraints on
the winds of coronal main sequence stars, thanks to an astrospheric
absorption signature created by the interaction between the stellar winds and
the interstellar medium.  We report the results of a new HST survey of M
dwarf stars, yielding six new detections of astrospheric absorption.  We
estimate mass-loss rates for these detections, and upper limits for
nondetections.  These new constraints allow us to characterize the nature of
M dwarf winds and their dependence on coronal activity for the first time.
For a clear majority of the M dwarfs, we find winds that are weaker or
comparable in strength to that of the Sun, i.e.\ $\dot{M}\leq 1$ $\dot{M}_{\odot}$.
However, two of the M dwarfs have much stronger winds:
YZ~CMi (M4 Ve; $\dot{M}=30$ $\dot{M}_{\odot}$)
and GJ~15AB (M2 V+M3.5 V; $\dot{M}=10$ $\dot{M}_{\odot}$).  Even these winds are
much weaker than expectations if the solar relation between flare energy
and coronal mass ejection (CME) mass extended to M dwarfs.  Thus, the solar
flare/CME relation does not appear to apply to M dwarfs, with important
ramifications for the habitability of exoplanets around M dwarfs.  There is
evidence for some increase in $\dot{M}$ with coronal activity as quantified
by X-ray flux, but with much scatter.  One or more other factors
must be involved in determining wind strength besides spectral type and
coronal activity, with magnetic topology being one clear possibility.

\end{abstract}


\section{Introduction}

     Most stars are known to exhibit some degree of mass loss, and for
many types of stars these stellar winds play an important role in
stellar evolution.  This is even true for the rather weak winds
of cool main sequence stars, which provide the means by which
these stars shed angular momentum, slowing their rotation with time
\citep{spm12,fg13,fg15,cpj15a,cpj15b,ja20}.
The solar wind, which is believed to be representative of this type of
wind, is a hot, fully ionized wind, with a relatively low mass loss
rate of $\dot{M}_{\odot}=2\times 10^{-14}$ M$_{\odot}$ yr$^{-1}$.
\citet{ajf19} provide estimates of the rate of angular
momentum loss for the Sun associated with this wind, based on direct
measurement.  Understanding how the solar rotation has evolved with
time is an active area of theoretical research \citep{aav21}, but this
requires observational constraints on the winds of solar-like stars
with various ages to constrain the solar wind evolution
\citep{tks13a,tks13b,vsa16,vr16,ms20}.

     The solar wind's existence is a natural consequence of the
heating processes that generate the hot solar corona ($T\sim 10^6$~K).
The exact nature of these processes is still uncertain, but coronal
heating is clearly associated with the conversion of magnetic energy
to thermal energy \citep{src12,src19}.
The solar corona and solar wind
are therefore linked to the solar dynamo that generates magnetic fields
in the Sun's outer convection zone that then emerge through the
photosphere, resulting in sunspots and other manifestations such as the
hot corona that represents the Sun's outermost atmospheric layer.

     Hot stellar coronae are most easily observed in X-rays, and X-ray
observations from missions such as {\em Einstein}, {\em ROSAT},
{\em Chandra}, and {\em XMM-Newton} have proven that X-ray emitting coronae
are ubiquitous for main sequence stars with spectral types later than
about A5~V \citep{js85,mg04}.  This makes
sense, as stars later than A5~V should all have outer convection zones
that can house stellar dynamos for generating the magnetic fields that
ultimately lead to stellar coronae and coronal X-ray emission.
In contrast, stars earlier than A5~V possess fully radiative interiors,
with no outer convection zone to host a dynamo.

     Given that all main sequence stars later than A5~V seem to have
hot coronae, it is generally assumed that all have coronal winds analogous
to the solar wind as well.  However, it has been difficult to
verify this, and to study how coronal winds vary with
stellar age, activity, and spectral type, because coronal winds are very
hard to detect and study observationally.  Such winds have densities
that are too low and they are too highly ionized to provide the kinds of
spectral diagnostics that are used to study more massive winds, such as
the radiation pressure driven winds of hot stars \citep{jp08} and
the warm winds of cool giants and supergiants
\citep{bew16,sh18,gr18}.

     Besides the solar wind, no coronal wind has ever been
directly detected to this day.  Attempts to detect radio free-free
emission from the ionized coronal winds have led only to nondetections.
Upper limits on mass-loss rates from these nondetections are orders of
magnitude stronger than the solar wind, indicating that radio
observations may not be sufficiently sensitive to provide many
useful diagnostics for the foreseeable future \citep{bf17}.
Another direct method to try to detect coronal winds is through
X-ray emission generated by charge exchange between the ionized wind
and interstellar neutrals that approach the star, but once again there
has not yet been a successful detection \citep{bjw02}.

     The only existing measurements of coronal winds rely on diagnostics
that detect the winds indirectly.  One such diagnostic is that of
slingshot prominences.  For certain very rapidly rotating stars, variable
H$\alpha$ absorption transients have been observed that are
believed to be indicative of material at chromospheric
temperatures trapped within very large prominences that lie far above
the stellar surface \citep{acc89}.  For at least
some of these stars, \citet{mj19} have recently argued
that the H$\alpha$ absorption can be used to estimate a plausible stellar
mass loss rate, the idea being that the absorption is sampling coronal wind
from the star that is temporarily trapped in the giant prominence,
subsequently cooling to chromospheric temperatures in the process.  The
material therefore is temporarily observable via H$\alpha$ absorption,
until the slingshot prominence erupts and releases the material
back into the stellar wind.  Mass-loss rates inferred from this diagnostic
are 2--4 orders of magnitude stronger than the solar wind.  This is
intuitive, since these rapidly rotating stars are much more active than
the Sun \citep{mj19}.

     Another indirect wind diagnostic is absorption from evaporating
exoplanetary atmospheres that transit in front of the star.  This
absorption has been detected in spectroscopic observations
of the H~I Lyman-$\alpha$ line by the {\em Hubble Space Telescope} (HST)
\citep{avm03,alde10,vb16}.  These observations are normally seen as being
of interest due to their diagnostic potential for studying
exoplanetary atmospheric structure and
evolution \citep[e.g.,][]{ae10,kgk14,ems16},
but they may be equally valuable for providing measurements of the
stellar wind, as the amount of absorption depends in part on the
density and velocity of the stellar wind at the exoplanet, as
demonstrated by \citet{aav17} for the M dwarf GJ~436.

     However, the indirect wind diagnostic that has provided the most
numerous wind measurements so far is astrospheric Lyman-$\alpha$
absorption, analogous to absorption that is also observable from
our own heliosphere.  In the immediate vicinity of the Sun,
hydrogen in the interstellar medium (ISM) is partly ionized and
partly neutral.  As the Sun moves through the ISM, the ISM plasma is
heated, compressed, and decelerated as it piles up outside the Sun's
heliopause, which is the contact surface separating the plasma flows
of the ISM and the solar wind.  Thanks to charge exchange, the
characteristics of the plasma outside the heliopause are transmitted
to the ISM neutrals as well, creating what has been called a ``hydrogen
wall'' outside the heliopause.  When \citet{jll96} analyzed the
Lyman-$\alpha$ emission observed from the two components of the very
nearby binary $\alpha$~Cen~AB (G2~V+K0~V), they found that ISM absorption
could not account for all of the absorption observed within the
chromospheric emission line.  The excess absorption could be explained
as a combination of absorption from the hydrogen wall around our own
heliosphere, and the analogous hydrogen wall around
$\alpha$~Cen~AB \citep{kgg97}.

     Stronger stellar winds will lead to larger astrospheres and thicker
hydrogen walls, with higher column densities and more Lyman-$\alpha$
absorption.  Thus, the astrospheric Lyman-$\alpha$ absorption can be
used as a diagnostic of stellar wind strength.  This technique has
been applied to numerous stars, including $\alpha$~Cen~AB
\citep{bew01,bew05a,bew05b,bew14,bew18}.
The astrospheric absorption diagnostic has provided more
coronal wind constraints than any other diagnostic, but there are still
only 16 astrospheric detections, leading to 16 published stellar mass
loss rate ($\dot{M}$) measurements, ranging from
$\dot{M}=0.15~\dot{M_{\odot}}$ for DK~UMa (G4~III-IV) to
$\dot{M}=100~\dot{M_{\odot}}$ for 70~Oph~AB (K0~V+K5~V).  The
astrospheric absorption signature implies that astrospheres might
in principle be detected in emission, from stellar Lyman-$\alpha$
photons scattered in the hydrogen walls.  However, the expected surface
brightness is very low, and an attempt made to detect this emission
was unsuccessful \citep{bew03}.

     This paper focuses on new astrospheric Lyman-$\alpha$ absorption
constraints for M dwarf stars, as there are only two prior astrospheric
detections for M dwarfs.  One is EV Lac (M3.5~Ve), with
$\dot{M}=1~\dot{M_{\odot}}$ \citep{bew05a}.  This is
a surprisingly modest wind considering that this is one of the most
notoriously active flare stars known, which might have been thought
to have a strong wind just due to coronal mass ejections (CMEs) associated
with the flaring \citep{jjd13}.  The second detection is
a very recent result for a much less active star, GJ~173 (M1~V), with
$\dot{M}=0.75~\dot{M_{\odot}}$ (Vannier et al., in preparation).  Our goal here
is to dramatically increase the number of wind constraints for M dwarfs
using new HST observations, enough to truly characterize the
winds of M dwarf stars for the first time, and assess how they vary
with coronal activity.

\section{New M Dwarf Observations from HST}

\begin{deluxetable}{cccccc}
\tabletypesize{\scriptsize}
\tablecaption{HST/STIS Observations}
\tablecolumns{6}
\tablewidth{0pt}
\tablehead{
  \colhead{Star} & \colhead{Spectral Type}&\colhead{Start Time}&
    \colhead{Grating}&\colhead{Wavelengths (\AA)} & \colhead{Exp. Time (s)}}
\startdata
GJ 15A & M2 V & 2019-07-03 17:45:15 & E230H & 2574--2851 & 1826 \\
       &      & 2019-07-03 19:17:37 & E140M & 1150--1700 & 2906 \\
       &      & 2019-07-08 15:30:42 & E140M & 1150--1700 & 5279 \\
GJ 205 &M1.5 V& 2019-08-07 00:47:52 & E230H & 2574--2851 & 1673 \\
       &      & 2019-08-07 02:03:49 & E140M & 1150--1700 & 2810 \\
       &      & 2019-09-15 10:05:13 & E140M & 1150--1700 & 5027 \\
GJ 273 &M3.5 V& 2019-09-20 03:12:37 & E230H & 2574--2851 & 1575 \\
       &      & 2019-09-20 04:26:55 & E140M & 1150--1700 & 2813 \\
       &      & 2019-09-23 23:11:13 & E140M & 1150--1700 & 4973 \\
YZ CMi & M4 V & 2019-10-15 21:18:05 & E230H & 2574--2851 & 1719 \\
       &      & 2019-10-15 22:35:09 & E140H & 1160--1360 & 2810 \\
       &      & 2019-10-18 15:53:33 & E140H & 1160--1360 & 5024 \\
GJ 338A& M0 V & 2019-01-09 13:31:04 & E230H & 2574--2851 & 1939 \\
       &      & 2019-01-09 14:42:00 & E140M & 1150--1700 & 3001 \\
       &      & 2019-01-10 16:22:32 & E140M & 1150--1700 & 5481 \\
GJ 588 &M2.5 V& 2018-08-23 09:34:20 & E230H & 2574--2851 & 1812 \\
       &      & 2018-08-23 10:49:35 & E140M & 1150--1700 & 3046 \\
       &      & 2018-09-15 05:30:18 & E140M & 1150--1700 & 5419 \\
GJ 644B&M4 V+M4 V&2018-08-26 13:51:02&E230H & 2574--2851 & 1782 \\
       &      & 2018-08-26 15:07:22 & E140H & 1160--1360 & 2953 \\
       &      & 2018-08-18 15:01:02 & E140H & 1160--1360 & 2337 \\
GJ 860A& M3 V & 2019-03-25 03:37:39 & E230H & 2574--2851 & 1879 \\
       &      & 2019-03-25 04:53:10 & E140M & 1150--1700 & 3057 \\
       &      & 2019-03-25 19:14:27 & E140M & 1150--1700 & 5521 \\
GJ 887 & M2 V & 2018-12-13 04:21:21 & E230H & 2574--2851 & 1866 \\
       &      & 2018-12-13 05:38:18 & E140M & 1150--1700 & 3008 \\
       &      & 2019-10-30 13:18:21 & E140M & 1150--1700 & 2347 \\
\enddata
\end{deluxetable}
     We here analyze the H~I Lyman-$\alpha$ absorption observed toward nine
nearby M dwarf stars observed by the Space Telescope Imaging
Spectrograph (STIS) instrument on board HST.  The observations are listed in
Table 1.  In choosing our target stars, we have focused on early M dwarfs,
which will have higher line fluxes than later type M
dwarfs, and will probably have stronger, more detectable winds simply due
to larger surface areas.  We use the Gliese-Jahreiss catalog numbers for the
star names in Table~1, with the exception of YZ~CMi (GJ~285), where we
use the variable star name by which that star is more commonly known.

     All of our chosen targets are within 7~pc.  The close
proximity not only increases the line fluxes, which maximizes the
signal-to-noise (S/N) of our spectra, but also greatly improves the
odds of detecting the astrospheric Lyman-$\alpha$ absorption
signature for other reasons that we now describe.
Within 7~pc, 10 out of 13 independent
lines of sight toward cool main sequence stars previously observed
by HST have yielded successful detections of astrospheric absorption,
but this detection fraction decreases dramatically as stellar distance
increases \citep{bew18}.

     The primary reason for the low astrospheric detection likelihood
beyond 7~pc is that the Sun lies within a fully ionized
region of the ISM called the Local Bubble (LB), which stretches
about 100~pc from the Sun in most directions \citep{jlv10,byw10,rl14}.
The fully ionized plasma
that predominates within the LB injects no neutrals into astrospheres
embedded within it, meaning no ``hydrogen wall'' structures, and
therefore no astrospheric absorption regardless of wind strength.
However, there are small clouds of cooler, partially neutral material
within the LB, and our Sun happens to lie within one of these.  This
is why we observe heliospheric Lyman-$\alpha$ absorption for some
lines of sight, and why for nearby stars we often see the analogous
astrospheric absorption from the ``hydrogen wall'' around the observed 
star.  But odds of successful astrospheric detection decrease quickly
beyond 7~pc, with all or most nondetections beyond this distance
likely being due to the ISM around the star being fully ionized.
One particularly
unfortunate implication of this is that such nondetections do
not even yield useful upper limits for mass-loss rates.  However,
this is not necessarily true within 7~pc, where the high astrospheric
detection fraction is consistent with the ISM within that distance
being partly neutral, meaning
that nondetections can be interpreted as being due to weak
stellar winds \citep{bew18}.

     We use spectra processed using the standard HST/STIS software,
available in the HST archive.  As indicated in Table~1, the primary HST/STIS
observation for each
target is an exposure of a spectral range containing the primary line of
interest, H~I Lyman-$\alpha$ at 1216~\AA.  In most cases, the E140M
grating is used to observe the 1150--1700~\AA\ region.  For the two most
active M dwarfs with the highest expected line fluxes, YZ~CMi and GJ~644B,
E140M was replaced by the higher resolution E140H grating, at the cost of
reducing the spectral coverage to 1160--1360~\AA.  For ease of scheduling,
the observations were separated into two separate visits, explaining the
two separate E140M (or E140H) observations listed for each target in Table~1.
We coadd the individual spectra into a single spectrum for our analysis.

     For each target, in addition to the far-UV (FUV) E140M/E140H spectrum we
also obtain a high resolution spectrum of the near-UV (NUV) 2574--2851~\AA\
region using the E230H grating, which contains the Mg~II h\&k lines at
2803 and 2796~\AA, respectively, and also Fe~II lines at 2600 and 2586~\AA.
There are two reasons these lines are of interest.  One is that the ISM
Mg~II and Fe~II absorption lines allow us to study the ISM velocity
structure, and the second is that the Mg~II chromospheric emission lines
are useful models for the shape of the intrinsic stellar H~I Lyman-$\alpha$
emission line.

     With regards to the first point, the Mg~II and Fe~II ISM lines are
very narrow, which is why they allow us to study the ISM velocity structure
along the line of sight to our observed stars.  This is not really
possible for the broader ISM H~I and D~I (deuterium) Lyman-$\alpha$
absorption, where the individual velocity components are completely blended.
Knowledge of the ISM velocity structure improves the accuracy of the analysis
of the H~I+D~I Lyman-$\alpha$ line, where the primary purpose is to separate
the ISM Lyman-$\alpha$ absorption from any heliospheric or astrospheric
absorption that might be present.

     With regards to the second point, our analysis of the H~I Lyman-$\alpha$
line requires the reconstruction of the stellar chromospheric H~I
Lyman-$\alpha$ emission line, which is highly absorbed by the very broad
ISM H~I absorption.  The Mg~II h\&k emission lines provide a useful model for
the intrinsic shape of the stellar H~I Lyman-$\alpha$ emission line, as
Mg~II h\&k and H~I Lyman-$\alpha$ are all highly opaque chromospheric lines,
with similar profiles in the solar spectrum.

     Some additional comment is necessary for the GJ~644B observation.
In the HST archive, the target is listed as GJ~644A, which is an M3~V star.
However, in the optical GJ~644A is nearly identical in brightness with
GJ~644B, which is a spectroscopic M4~V+M4~V binary.  Furthermore, the A and
B components of this system are close enough ($\sim 0.2^{\prime\prime}$
separation) that both are in the target acquisition image.  The B component
ended up slightly brighter than the A component in this image, so it is the B
component that was acquired and observed by HST.  Whether the A or B
component is observed is unimportant for our purposes, as the stars are close
enough to reside within the same astrosphere, and therefore the astrospheric
absorption should be identical toward both (see section~4.2).

\section{Absorption Line Analysis}

\subsection{The Mg~II and Fe~II Lines}

     Although our focus is on searching for and measuring astrospheric
H~I Lyman-$\alpha$ absorption, this absorption is
highly blended with ISM absorption, so our analysis necessarily involves
measurements of ISM absorption along the line of sight (LOS) to each of our
target stars.  These measurements are naturally of interest in their own
right, providing new information about the character of the ISM in the
solar neighborhood.  We first concentrate on ISM absorption lines of Mg~II
and Fe~II in the E230H spectra.  The narrowness of these lines is useful
for revealing the velocity structure of the ISM toward
the star, for consideration in the following Lyman-$\alpha$ analysis.  

\begin{figure}[t]
\plotfiddle{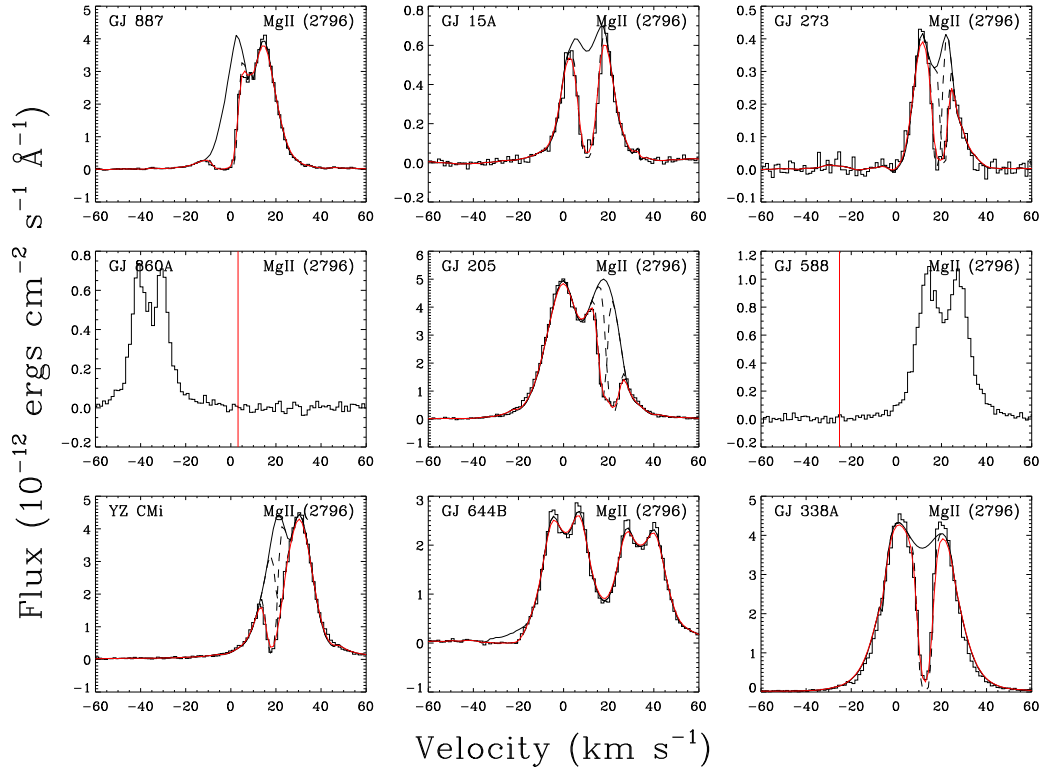}{3.5in}{90}{55}{55}{210}{-25}
\caption{HST/STIS E230H spectra of the Mg~II k lines of
  our nine M dwarf target stars (at 2796~\AA), plotted on a
  heliocentric velocity scale.  Fits to the ISM absorption
  lines seen within the chromospheric emission lines are
  shown in red, with two ISM components inferred for the
  GJ~273, GJ~205, and YZ~CMi lines of sight (dashed lines).
  No ISM absorption is seen toward GJ~860A or GJ~588, with
  the expected location of the absorption (vertical red line)
  too far from the emission line.}
\end{figure}
     Figure~1 shows the Mg~II k lines of our nine M dwarf target stars,
plotted on a heliocentric velocity scale.  The chromospheric emission lines
all show a self-reversal near line center, which is typical for earlier
type stars as well.  Two separate Mg~II lines are seen for GJ~644B, as this
is a 2.97~day spectroscopic binary \citep{ds00,tm01}
with the binary observed near quadrature, where the
velocity difference between GJ~644Ba (M4~V) and GJ~644Bb (M4~V) is large
enough to almost completely separate the two lines.

     The stellar Mg~II lines are all very narrow, compared to those observed
from other types of cool stars.  This is expected, since M dwarfs are
the least luminous of main sequence stars, and for high opacity
chromospheric resonance lines like Mg~II h\&k there is a known relation
between line width and stellar absolute magnitude \citep[e.g.,][]{ac01}.
This relation is generally referred to as the
Wilson-Bappu effect \citep{ocw57}.  The narrowness of the lines
complicates the analysis of the ISM absorption lines in various ways.

     For most of the stars in Figure~1, narrow ISM absorption is superposed
on the stellar Mg~II emission, but in a couple cases the narrowness of the
Mg~II emission means that the ISM absorption misses the lines entirely.
For these two stars, GJ~860A and GJ~588, Figure~1 shows the expected
location of the absorption based on the Local Interstellar Cloud (LIC) flow
vector of \citet{sr08}.  Unlike earlier type main sequence stars,
M dwarfs provide no NUV continuum emission to provide background flux for
the ISM absorption lines, if they miss the chromospheric Mg~II emission.
Even for the other cases where the ISM absorption lands within the emission
line, the narrowness of the emission causes problems with inferring the
shape of the emission background for the absorption.  For example,
it makes it harder to separate the ISM absorption from the self-reversal
of the Mg~II emission, when the ISM absorption is near line center
(e.g., GJ~15A, GJ~273, GJ~338A).

\begin{deluxetable}{llccccc}
\tabletypesize{\scriptsize}
\tablecaption{Absorption Line Fit Parameters\tablenotemark{a}}
\tablecolumns{7}
\tablewidth{0pt}
\tablehead{
  \colhead{Star}&\colhead{Ion}&\colhead{$\lambda_{rest}$\tablenotemark{b}} &
    \colhead{ISM} & \colhead{$v$\tablenotemark{c}} & \colhead{$b$} &
    \colhead{log N} \\ 
  \colhead{}&\colhead{}&\colhead{(\AA)} & \colhead{Cloud}&\colhead{(km~s$^{-1}$)} &
    \colhead{(km~s$^{-1}$)} & \colhead{log(cm$^{-2}$)}} 
\startdata
GJ 15A & Mg II & 2796.3543, 2803.5315 & LIC & $10.91\pm 0.32$ &
   $3.50\pm 0.23$ & $12.63\pm 0.02$ \\
       & H I (ISM)&1215.6682, 1215.6736&LIC &  $8.85\pm 0.46$ &
   $12.13\pm 0.28$ & $18.02\pm 0.02$ \\
       &H I (HS/AS)&1215.6682, 1215.6736&...&  $-10.6\pm 6.6$ &
  $14.3\pm 1.9$ & $14.98\pm 0.67$ \\
GJ 205 & Mg II & 2796.3543, 2803.5315 & ?   & $17.31\pm 0.19$ &
   $1.98\pm 0.27$ & $12.13^{+0.10}_{-0.11}$ \\
       & Mg II & 2796.3543, 2803.5315 & LIC & $21.93\pm 0.16$ &
   $2.92\pm 0.16$ & $12.49\pm 0.05$ \\
       & H I (ISM)&1215.6682, 1215.6736& ?  & $17.57\pm 0.22$ &
   $8.98\pm 0.56$ & $17.24\pm 0.01$ \\
       & H I (ISM)&1215.6682, 1215.6736&LIC & ($22.19\pm 0.22$) &
   ($8.98\pm 0.56$) & ($17.60\pm 0.01$) \\
       &H I (HS/AS)&1215.6682, 1215.6736&...& $19.76\pm 0.18$ &
  $17.00\pm 0.69$ & $15.70\pm 0.18$ \\
GJ 273 & Mg II & 2796.3543, 2803.5315 & LIC & $18.28\pm 0.92$ &
   $1.66\pm 0.44$ & $12.41^{+0.10}_{-0.12}$ \\
       & Mg II & 2796.3543, 2803.5315 & Aur & $21.38\pm 0.40$ &
   $1.33\pm 0.64$ & $12.33\pm 0.05$ \\
       & H I (ISM)&1215.6682, 1215.6736&LIC & $18.25\pm 0.19$ &
   $11.94\pm 0.14$ & $17.86\pm 0.01$ \\
       & H I (ISM)&1215.6682, 1215.6736&Aur & ($21.36\pm 0.19$) &
   ($11.94\pm 0.14$) & ($17.78\pm 0.01$) \\
YZ CMi & Mg II & 2796.3543, 2803.5315 & LIC & $18.11\pm 0.31$ &
   $2.25\pm 0.26$ & $12.44^{+0.11}_{-0.14}$ \\
       & Mg II & 2796.3543, 2803.5315 & Aur & $21.73\pm 0.53$ &
   $2.72\pm 0.43$ & $12.00^{+0.10}_{-0.12}$ \\
       & Fe II & 2586.6500, 2600.1729 & LIC & $16.49\pm 0.53$ &
   $2.31\pm 0.66$ & $12.42^{+0.12}_{-0.16}$  \\
       & Fe II & 2586.6500, 2600.1729 & Aur & $21.16\pm 0.70$ &
   $2.67\pm 0.70$ & $12.20^{+0.11}_{-0.14}$  \\
       & H I (ISM)&1215.6682, 1215.6736&LIC & $18.43\pm 0.15$ &
   $12.15\pm 0.13$ & $17.89\pm 0.01$ \\
       & H I (ISM)&1215.6682, 1215.6736&Aur & ($22.06\pm 0.15$) &
   ($12.15\pm 0.13$) & ($17.45\pm 0.01$) \\
       &H I (HS/AS)&1215.6682, 1215.6736&...& $11.8\pm 0.5$ &
  $29.4\pm 0.6$ & $14.59\pm 0.04$ \\
GJ 338A& Mg II & 2796.3543, 2803.5315 & LIC & $12.58\pm 0.09$ &
   $2.28\pm 0.62$ & $12.48^{+0.19}_{-0.33}$ \\
       & Fe II & 2586.6500, 2600.1729 & LIC & $12.23\pm 0.43$ &
   $2.20\pm 0.37$ & $12.33\pm 0.03$ \\
       & H I (ISM)&1215.6682, 1215.6736&LIC & $12.11\pm 0.17$ &
   $10.67\pm 0.13$ & $17.97\pm 0.01$ \\
       &H I (HS/AS)&1215.6682, 1215.6736&...&  $6.4\pm 4.5$ &
  $18.6\pm 2.0$ & $14.76\pm 0.29$ \\
GJ 588 & H I (ISM)&1215.6682, 1215.6736&  G & $-26.58\pm 0.56$ &
   $11.06\pm 0.50$ & $18.12\pm 0.01$ \\
       &H I (HS/AS)&1215.6682, 1215.6736&...& $-1.4\pm 6.3$ &
  $17.6\pm 2.2$ & $13.79\pm 0.32$ \\
GJ 644B& Mg II & 2796.3543, 2803.5315 & Mic?& $-26.10\pm 0.59$ &
   $4.8\pm 2.2$ & $13.30^{+0.92}_{-0.22}$  \\
       & H I (ISM)&1215.6682, 1215.6736&Mic?& $-25.45\pm 0.13$ &
   $13.75\pm 0.10$ & $18.40\pm 0.01$ \\
GJ 860A& H I (ISM)&1215.6682, 1215.6736&Eri & $-0.20\pm 0.36$ &
   $7.64\pm 0.97$ & $17.78\pm 0.02$ \\
       &H I (HS/AS)&1215.6682, 1215.6736&...& $-1.69\pm 0.40$ &
  $19.6\pm 1.3$ & $15.60\pm 0.30$ \\
GJ 887 & Mg II & 2796.3543, 2803.5315 & LIC & $-2.74\pm 0.21$ &
   $3.45\pm 0.06$ & $13.33\pm 0.05$ \\
       & Fe II & 2586.6500, 2600.1729 & LIC & $-4.00\pm 0.87$ &
   $3.13\pm 0.75$ & $12.60^{+0.15}_{-0.24}$  \\
       & H I (ISM)&1215.6682, 1215.6736&LIC & $-3.41\pm 0.16$ &
   $12.20\pm 0.12$ & $18.10\pm 0.01$ \\
       &H I (HS/AS)&1215.6682, 1215.6736&...& $-17.1\pm 5.8$ &
  $33.7\pm 2.2$ & $14.15\pm 0.15$ \\
\enddata
\tablenotetext{a}{Values in parentheses are fixed relative to other component
  (see text).}
\tablenotetext{b}{Rest wavelengths of measured lines, in vacuum.}
\tablenotetext{c}{Central velocity in a heliocentric rest frame.}
\end{deluxetable}
     We fit the observed ISM absorption lines using procedures used in
many past studies \citep{sr02,sr04}, and described more
briefly below.  Both the h and k lines
are fitted simultaneously, though Figure~1 shows only the stronger k line.
The best fit is determined by $\chi^2$ minimization \citep{prb92}.
The fits include corrections for
instrumental broadening \citep{sh12}.  For three lines of sight
(GJ~273, GJ~205, YZ~CMi), fitting the data requires two separate ISM
components, as shown in Figure~1.  Each absorption component is defined by
three paramters; the central velocity ($v$), Doppler broadening
parameter ($b$), and column density ($N$).  The parameters resulting from our
fits are listed in Table~2.  Note that the quoted 1$\sigma$ uncertainties
only include random errors induced by the noise in the data, and do not
include systematic errors such as uncertainties in the shape of the
background line profile, which likely dominates the uncertainties
in the analysis.  For each detected ISM component, we have used the
cloud radial velocities and local ISM maps from \citet{sr08}
to identify likely nearby clouds responsible for the absorption, and
these are listed in the fourth column of Table~2.  Absorption from the
LIC, the cloud in which the Sun resides, is seen in all but three
directions (GJ~588, GJ~644B, and GJ~860A).  The reason it is not seen in
three cases is presumably that we are close to the edge of the LIC in those
directions.  For GJ~588 and GJ~644B, this is unsurprising based
on the shape of the LIC inferred by \citet{jll19}.  However, the lack of
LIC absorption toward GJ~860A is unexpected, suggesting that the LIC model
may require revision in that direction.

     The ISM absorption toward GJ~644B nearly misses the emission, lying
in the far blue wing of the line.  The measured Doppler parameter of the
single ISM component fitted to this absorption, $b=4.8\pm 2.2$ km~s$^{-1}$,
is suspiciously high, suggesting that multiple ISM components are probably
present, but the low S/N this far in the wing of the line does
not allow us to confirm the presence of multiple components.

     There are two Fe~II lines in our E230H spectra that can in principle
provide further ISM absorption diagnostics, with rest wavelengths of
2586.6500~\AA\ and 2600.1729~\AA.  These lines can be measured in the
same way as Mg~II h\&k.  However, the stellar Fe~II emission lines
are significantly narrower and weaker than the Mg~II lines.  Thus, in most
cases the ISM lines either miss the stellar emission entirely, or are observed
with insufficient S/N for useful measurements to be made.  We report Fe~II
measurements for only YZ~CMi, GJ~338A, and GJ~887 in Table~2.

\subsection{The H~I Lyman-$\alpha$ Line}

\begin{figure}[t]
\plotfiddle{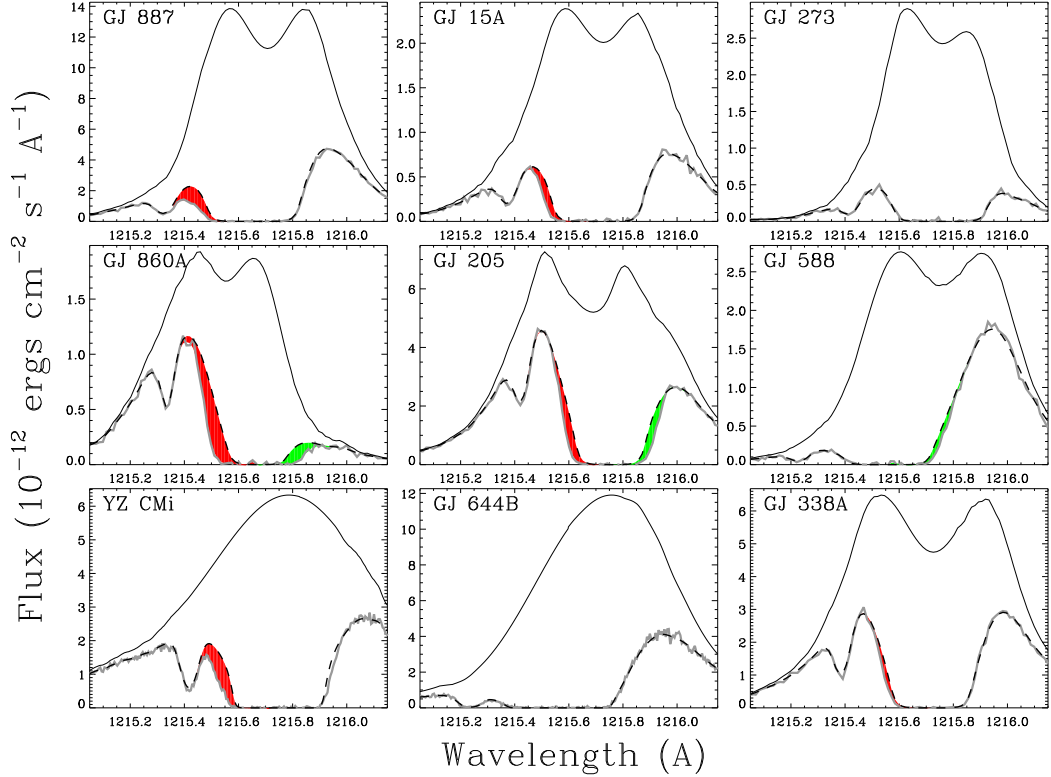}{3.5in}{90}{60}{60}{220}{-50}
\caption{HST/STIS spectra of the H~I Lyman-$\alpha$ lines of
  our nine M dwarf target stars.  The chromospheric emission
  lines are obscured by very broad H~I absorption, and narrow
  D~I absorption $-0.33$~\AA\ blueward of the H~I absorption.
  Fits to the absorption are performed, which involves
  reconstruction of the stellar line profile over the
  absorption.  The dashed lines indicate the ISM absorption
  alone inferred for each LOS, assuming self-consistency
  between H~I and D~I.  In only two cases (GJ~273 and
  GJ~644B) does this fit the data.  The other cases all show
  excess H~I absorption, either heliospheric absorption on the
  long wavelength side of the line (shown in green) and/or astrospheric
  absorption on the short wavelength side of the line (shown in red).}
\end{figure}
     The H~I Lyman-$\alpha$ lines of our nine target stars are
shown in Figure~2.  For each profile, the chromospheric emission
line is greatly obscured by absorption.  This absorption consists
of very broad, fully saturated absorption from H~I in between
the star and Earth, and much narrower absorption from deuterium
(D~I), which is $-0.33$~\AA\ blueward of the H~I absorption.  The D~I
absorption is entirely interstellar, but the H~I absorption can
include contributions from heliospheric and/or astrospheric
absorption.  It should be noted that narrow geocoronal
Lyman-$\alpha$ emission has been removed from the spectra.  This
removal can be problematic if the geocoronal emission is blended with
any of the emission observed from the star \citep{bew05b}.
However, for all our observations the geocoronal emission is a
weak, narrow emission line completely within the saturated core
of the ISM absorption, unblended with any of the stellar emission and
therefore easy to remove.

     Hydrogen is the only element abundant enough for the
heliospheric and/or astrospheric (HS/AS) absorption signature to be
detectable.  In the analysis of the Lyman-$\alpha$ lines, the
D~I absorption is crucial, as the presence of HS/AS
absorption reveals itself through discrepancies between
the H~I and D~I absorption profiles.  Our analysis mirrors many past
studies, and is described most extensively in \citet{bew05b}.
The H~I and D~I absorption lines are fitted simultaneously using
$\chi^2$ minimization, taking into
account the two hyperfine components of the H~I and D~I lines.
We force the H~I and D~I absorption features to have
self-consistent central velocities [e.g., $v({\rm HI})=v({\rm DI})$]
and Doppler parameters.  Given that the H~I and D~I lines are
dominated by thermal broadening, the latter constraint means
that $b({\rm HI})=\sqrt{2}\times b({\rm DI})$.  Finally, the
{\rm D/H} ratio has been found to be constant within the LB, with
a value of ${\rm D/H}=1.56\times 10^{-5}$ \citep{bew04},
so in the fits we force the column density ratio of D~I to H~I
to be consistent with this value.  In this way, all three of the
H~I parameters are actually related to D~I, meaning that in a
single-component fit to the H~I+D~I absorption there are
actually only three free parameters.

     In the few cases where the Mg~II analysis has indicated
multiple ISM components (see Table~2), we include these components
in the Lyman-$\alpha$ analysis, though the components are
hopelessly blended in the broader H~I and D~I lines.  In the
H~I+D~I fits, the velocity separation of the components is forced
to be consistent with the Mg~II fit.  For simplicity, we also
simply force the components to have the same column density
ratio as in the Mg~II fit, and we assume that the
components have identical Doppler parameters.  Parameters fixed
in this fashion are identified with parentheses in Table~2.

     The Lyman-$\alpha$ analysis not only involves fitting the
absorption, but also reconstructing the intrinsic stellar
Lyman-$\alpha$ profile in the process, which is not trivial
considering the broad extent of the H~I absorption.  In the
$\log N({\rm HI})\approx 18.0$ column density regime represented
by our nearby target stars, the Lorentzian natural line opacity
profile is relevant as well as the Gaussian thermal Doppler core.
The full opacity profile is a Voigt profile, which is a
convolution of the Lorentzian and Gaussian components.  The wing
absorption apparent far from line center in Figure~2
is associated with the former, while the Doppler core of the
opacity profile affects the width of the saturated
core of the absorption.

     In the analysis, an initial guess is
made for the background shape of the stellar Lyman-$\alpha$
line, with guidance from the profile of the Mg~II lines, which
like Lyman-$\alpha$ are highly opaque chromospheric resonance
lines that have similar profiles in the solar spectrum.  After
an initial fit is made to the data, the residuals of the fit
are then used to modify the assumed stellar line profile to
improve the fit.  Further iterations are made, if necessary.
The self-reversals inferred from Mg~II are generally preserved
in this process, but in practice the absorption fit parameters
are little affected by the exact shape of the stellar profile
near line center.

\begin{figure}[t]
\plotfiddle{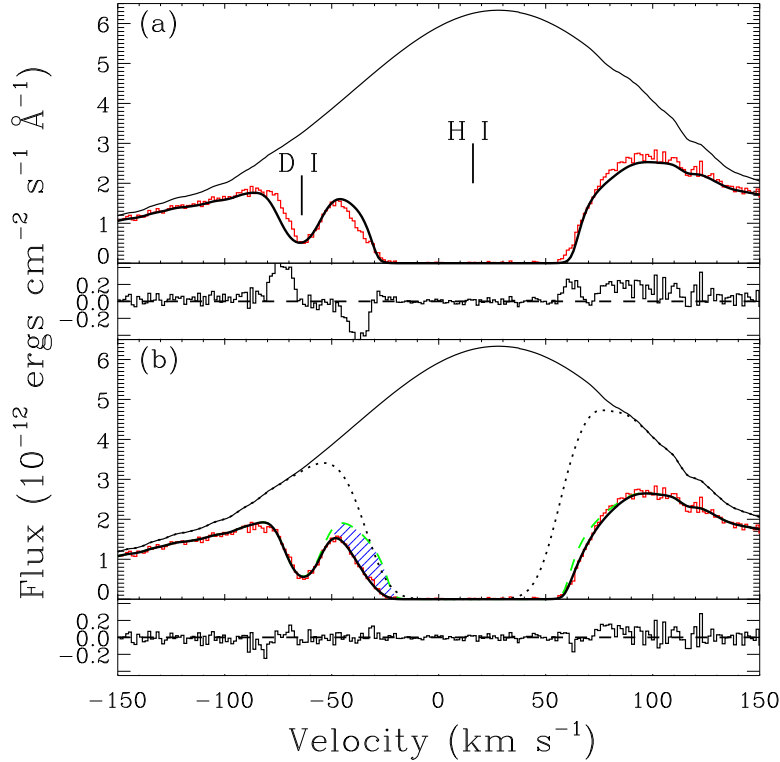}{3.5in}{0}{80}{80}{-240}{-295}
\caption{(a) A fit to the H~I+D~I Lyman-$\alpha$ line of
  YZ~CMi, assuming only ISM absorption is present, with
  residuals shown below the fit.  The
  H~I and D~I absorption are forced to be self-consistent,
  and the resulting fit is poor.  (b) A two-component fit
  to the Lyman-$\alpha$ line, representing absorption from
  both the ISM (green dashed line) and from the stellar
  astrosphere (black dotted line).  The combination of
  the two components (thick black line) fits the data.
  The hatched region explicitly indicates the excess
  absorption due to the astrosphere.}
\end{figure}
     Every effort is made to find an acceptable fit to the
data assuming only ISM absorption.  However,
we are successful in only two of the nine cases, GJ~273 and
GJ~644B.  For the other stars, the ISM-only fits are poor,
presumably due to the presence of HS/AS absorption
contributing to the H~I absorption but not D~I.  Using YZ~CMi
as an example, Figure~3(a) explicitly shows the best ISM-only
fit to the data, which is clearly unacceptable.  The fitted
D~I absorption is too broad.  This implies that the widths of
the H~I and D~I absorption are inconsistent, with H~I too broad
relative to D~I.  In other words, there must be excess H~I
absorption beyond that from the ISM, but not D~I because of the
much lower column density in this line.
Furthermore, the fitted D~I absorption profile is blueshifted
relative to the observed absorption, suggesting that the
excess H~I absorption is primarily on the blue side of the line,
which is indicative of an astrospheric absorption detection.

     The heliospheric and astrospheric absorption
produce excess absorption on opposite sides of the ISM
absorption, allowing us to the distinguish between the two.
When detectable, the heliospheric absorption
is apparent as excess absorption on the right side of the
ISM absorption.  This redshift is due to the deceleration and
deflection of ISM material as it approaches the heliopause.
In contrast, the astrospheric absorption is blueshifted relative
to the ISM absorption for the same reason, with the difference
being due to our viewpoint outside the astrosphere instead
of inside.

     For the seven cases where HS/AS absorption is present,
it is necessary to estimate the nature of the excess H~I
absorption with another fit.  For this purpose, we
add another absorption component to the fit that is designed
to approximate the HS/AS absorption regardless of whether
the excess absorption is heliospheric, astrospheric, or some
combination of the two.  Figure~3(b) shows an example of the
resulting fit for YZ~CMi, consistent with excess absorption
on the blue side of the line, which is the astrospheric
signature.  There is a hint of excess on the red side of the
line as well, which could be a sign of heliospheric absorption,
but we do not consider the excess sufficiently convincing to
deem this a heliospheric detection.

     As shown in Figure~2, we find evidence for
astrospheric signatures in six of the spectra (GJ~887, GJ~15A,
GJ~860A, GJ~205, YZ~CMi, and GJ~338A) and heliospheric
signatures in only three cases (GJ~860A, GJ~205, and GJ~588).
The detectability of heliospheric absorption depends primarily
on two factors.  The first is the ISM H~I column density, which
if high enough will broaden the ISM absorption enough to obscure
the heliospheric signature.  The second is the angle of the LOS
relative to the ISM flow direction seen by the Sun, with the 
heliospheric absorption much easier to detect closer to upwind
directions \citep{bew05b}.  For the three heliospheric
detections, we verify that the inferred heliospheric absorption
is consistent with expectations by comparing the
observed excess absorption with that predicted by a
heliospheric model from \citet{bew00}, which we have
found in past analyses to be successful in reproducing observed
heliospheric Lyman-$\alpha$ absorption.  We will say little
more about the heliospheric absorption, as we are here
more interested in the astrospheric detections and what they
imply about the winds of their associated stars.

     The final H~I Lyman-$\alpha$ fit parameters are listed
in Table~2.  We include the parameters of the HS/AS components,
although the properties of the heliosphere and/or astrosphere cannot
be successfully quantified using fits of this nature.
Hydrodynamic models are required for any meaningful quantitative
analysis of the HS/AS absorption, as will be described in
the next section.

     Finally, we note in passing that the intrinsic stellar
Lyman-$\alpha$ profiles are also useful products
of this analysis.  From such profiles, integrated
Lyman-$\alpha$ line fluxes can be measured, which are
excellent diagnostics of stellar chromospheric activity.
The Lyman-$\alpha$ line fluxes from our analysis
are listed and discussed by \citet{km20} and \citet{jll20}.

\section{Analysis of M Dwarf Astrospheric Absorption}

\subsection{Summary of M Dwarf Wind Constraints from HST}

\begin{deluxetable}{cllcccccc}
\tabletypesize{\scriptsize}
\tablecaption{Mass Loss Measurements for Coronal Winds}
\tablecolumns{9}
\tablewidth{0pt}
\tablehead{
  \colhead{ID \#} & \colhead{Star}&\colhead{Spectral}&\colhead{$d$} &
    \colhead{$V_{ISM}$} & \colhead{$\theta$} & \colhead{$\dot{M}$} &
    \colhead{log L$_{x}$} & \colhead{Radius} \\
  \colhead{}&\colhead{}&\colhead{Type}&\colhead{(pc)} &
    \colhead{(km~s$^{-1}$)}&\colhead{(deg)}&\colhead{($\dot{M}_{\odot}$)}&
    \colhead{} & \colhead{(R$_{\odot}$)} }
\startdata
\multicolumn{9}{l}{\underline{M DWARFS}} \\
 1 & Prox Cen      & M5.5 V      & 1.30 & 25 & 79 &$<0.2$& 27.22 & 0.14 \\
 2 &GJ 699\tablenotemark{a}& M4 V& 1.83 &121 & 43 &$<0.2$& 25.85 & 0.19 \\
 3 &GJ 411\tablenotemark{a}& M2 V& 2.55 &110 & 36 &$<0.1$& 26.89 & 0.36 \\
 4 & GJ 729        & M3.5 V      & 2.98 & 11 &178 & ...  & 27.06 & 0.20 \\
 5 &GJ 887\tablenotemark{a}& M2 V& 3.29 & 85 & 99 & 0.5  & 27.03 & 0.47 \\
 6 &GJ 15AB\tablenotemark{a}&M2 V+M3.5 V&3.56 & 28 & 95 & 10   & 27.37 &0.37+0.17 \\
 7 &GJ 273\tablenotemark{a}&M3.5 V&3.80 & 75 & 89 &$<0.2$& 26.54 & 0.31 \\
 8 &GJ 860AB\tablenotemark{a}&M3 V+M4 V& 4.01 & 47 & 55 & 0.15 & 27.72 &0.29+0.26 \\
 9 & AD Leo        & M4 V        & 4.97 & 13 &114 & ...  & 28.80 & 0.39 \\
10 & EV Lac        & M3.5 V      & 5.05 & 45 & 84 &  1   & 28.99 & 0.32 \\
11 &GJ 205\tablenotemark{a}&M1.5 V& 5.70& 70 & 79 & 0.3  & 27.66 & 0.59 \\
12 &GJ 588\tablenotemark{a}&M2.5 V&5.92 & 47 &139 & $<5$ & 27.00 & 0.43 \\
13 &YZ CMi\tablenotemark{a}& M4 V& 5.99 & 20 &114 & 30   & 28.57 & 0.33 \\
14&GJ 644AB\tablenotemark{a}& M3+M4+M4 V& 6.20 & 53 &136 & $<5$ & 29.04&0.3+0.3+0.3 \\
15&GJ 338AB\tablenotemark{a}&M0 V+M0 V  & 6.33 & 29 & 88 & 0.5  & 27.92 &0.60+0.60 \\
16 &GJ 436\tablenotemark{b}& M3 V& 9.75 & 79 & 97 &0.059 &26.76 & 0.40 \\
17 & GJ 173        & M1 V        & 11.2 & 38 & 43 & 0.75 & 26.84 & 0.42 \\
\multicolumn{9}{l}{\underline{GK STARS}} \\
18 & $\alpha$ Cen AB&G2 V+K0 V   & 1.35 & 25&79&0.46+1.54&26.99+27.32&1.22+0.86 \\
19 & $\epsilon$ Eri& K2 V        & 3.22 & 27 & 76 & 30   & 28.31 & 0.74 \\
20 & 61 Cyg A      & K5 V        & 3.48 & 86 & 46 & 0.5  & 27.03 & 0.67 \\
21 & $\epsilon$ Ind& K5 V        & 3.63 & 68 & 64 & 0.5  & 27.39 & 0.73 \\
22 & $\tau$~Cet    & G8 V        & 3.65 & 56 & 59 &$<0.1$& 26.69 & 0.77 \\
23 & 70 Oph AB     & K0 V+K5 V   & 5.09&37&120 &55.7+44.3&28.09+27.97&0.83+0.67 \\
24 & 36 Oph AB     & K1 V+K1 V   & 5.99 & 40&134&8.5+6.5 &28.02+27.89&0.69+0.59 \\
25 & $\delta$~Pav  & G8 IV       & 6.11 & 29 & 72 & 10   & 27.29 & 1.22 \\
26 & GJ~892        & K3 V        & 6.55 & 49 & 60 & 0.5  & 26.85 & 0.78 \\
27 & $\xi$ Boo AB  & G8 V+K4 V   & 6.70 & 32&131&0.5+4.5 &28.91+28.08&0.86+0.61 \\
28 & 61 Vir        & G5 V        & 8.53 & 51 & 98 & 0.3  & 26.87 & 0.99 \\
29 & $\delta$~Eri  & K0 IV       & 9.04 & 37 & 41 &  4   & 27.05 & 2.58 \\
30 & $\pi^1$ UMa   & G1.5 V      & 14.4 & 43 & 34 & 0.5  & 28.99 & 0.97 \\
31 & $\lambda$ And & G8 IV-III   &25.8 & 53 & 89 &  5   & 30.82 & 7.40 \\
32 & DK UMa        & G4 III-IV   & 32.4 & 43 & 32 & 0.15 & 30.36 & 4.40 \\
\multicolumn{9}{l}{\underline{SLINGSHOT PROMINENCE STARS}} \\
33 & V374 Peg      & M3.5 V      &  9.1 &  9 & 85 &  200 & 28.45 & 0.34 \\
34 & AB Dor        & K0 V        & 15.3 & 31 &157 &  350 & 29.87 & 0.93 \\
35 & HK Aqr        & M0 V        & 24.9 & 14 &110 &   50 & 29.01 & 0.59 \\
36 & Speedy Mic    & K3 V        & 66.7 & 25 &109 &  130 & 31.06 & 1.06 \\
37 & LQ Lup        & G8 IV       &146.7 & 24 &165 & 4500 & 30.90 & 1.30 \\
\enddata
\tablenotetext{a}{New analysis.}
\tablenotetext{b}{Exoplanet transit absorption.}
\end{deluxetable}
     The central goal of our project is to characterize the
winds of M dwarf stars using all available constraints.  To
that end, in the top section of Table~3 we have compiled a list
of all M dwarfs with relevant Lyman-$\alpha$ observations from
HST that can at least in principle yield wind constraints.
We include stars within 7~pc that are nondetections, since
within 7~pc nondetections can possibly provide upper limits
for $\dot{M}$ (see Section~2).  Stars listed include the ones
mentioned in Section~1, namely the astrospheric absorption
detections for EV~Lac and
GJ~173 \citep[][Vannier et al., in preparation]{bew05a}, and the transiting
exoplanet absorption constraint for GJ~436 \citep{aav17}.
Also listed is the upper limit for Proxima~Cen from
\citet{bew01}.  Another astrospheric nondetection, AD~Leo,
is also listed \citep{bew05b}, and we here assess whether
these data can provide a useful $\dot{M}$ upper limit.

     The nine target stars of our new HST observing program
account for most of the rest of the M dwarfs in Table~3.
The exceptions are three additional very nearby stars.  These
include GJ~729, which is part of the Mega-MUSCLES project
\citep[HST program GO-15071:][]{km20,jll20}.
The other two are GJ~699 and GJ~411, the analysis of which will
be described elsewhere (Youngblood et al., in preparation).  They
and GJ~729 are all astrospheric nondetections, and like the AD~Leo
case noted above we here try to infer upper limits for $\dot{M}$.
The second section of Table~3 lists
other $\dot{M}$ measurements from astrospheric Lyman-$\alpha$
constraints, as previously compiled by \citet{bew18}.  Finally, the
bottom section lists slingshot prominence stars with $\dot{M}$
measurements, from \citet{mj19}.

     In the case of binary stars, the companion stars are listed
if they are nearby enough that both stars will lie within the
same astrosphere, meaning that the astrospheric absorption will be
a diagnostic of the combined winds of both stars.  For example,
the GJ~860AB binary (M3~V+M4~V) has a separation of only
$2.4^{\prime\prime}$, corresponding to a plane-of-sky distance of 
only 9.6~au, easily close enough for both stars to reside within the
same astrosphere.  For GJ~644AB, there are actually three stars
involved, with the observed spectroscopic binary
GJ~644B (M4~V+M4~V) separated from GJ~644A (M3~V) by
$0.23^{\prime\prime}$, corresponding to a distance of only 1.5~au.

     The $V_{ISM}$ quantity in Table~3 is the ISM flow speed seen
by the star in stellar rest frame, and $\theta$ is the angle betwen
the upwind direction of the ISM flow and our LOS to the star.
These are quantities that must be known
to infer $\dot{M}$ from an astrospheric absorption detection, as
will become clear in Section~4.2.  Computing $V_{ISM}$ and $\theta$
requires knowledge of the stellar radial velocity and proper motion
(very well known for such nearby stars, and
obtainable from the SIMBAD database), and the LIC flow vector from
\citet{sr08}.  Given that multiple ISM velocity
components are often seen toward even very nearby stars, other
ISM flow vectors besides that of the LIC clearly exist near the Sun,
and may exist around some of our observed stars, particularly the
ones with multiple Mg~II components (see Figure~1).  However, within
7~pc the multiple ISM components, when present, are generally
closely spaced in velocity, meaning that the different ISM
vectors are similar.  Thus, our universal assumption of the LIC
vector should be a decent approximation.  As an example, the
existence of Aur cloud absorption toward GJ~273 means that GJ~273
probably lies within the Aur cloud instead of the LIC.  Using the
Aur cloud vector from \citet{sr08}, we estimate
$V_{ISM}=65$ km~s$^{-1}$ and $\theta=86^{\circ}$ for GJ~273,
representing only a small change from the $V_{ISM}=75$ km~s$^{-1}$ and
$\theta=89^{\circ}$ values computed using the LIC vector.

     The $\dot{M}$ column in Table~3 lists inferred mass loss
rates, with new results being described below in Section~4.2.
For four of the previously studied binaries, the $\dot{M}$
has been divided between the two stars, as described by \citet{bew18}.
We will be seeking to relate $\dot{M}$ to coronal
activity, so coronal X-ray luminosities (in ergs~s$^{-1}$) are
also listed, based mostly on {\em ROSAT} all-sky survey
measurements \citep{js04}.  For binaries that
have been resolved in X-rays, we list separate $\log L_X$
values for both stars.

     Finally, when comparing $\dot{M}$ and $\log L_X$ for
stars of different sizes, it is appropriate to
normalize by stellar surface area, so we also list in
Table~3 the stellar radii that we are assuming.  For
the M dwarfs, the radii are mostly from \citet{erh19}.
However, for GJ~644AB no individual radii
measurements exist for the three individual stars in this
system, so we simply assume a typical radius for mid-M dwarfs,
$R=0.3$ R$_{\odot}$, for all three stars.

\subsection{Hydrodynamic Modeling of Astrospheres}

     Our HST survey of nearby M dwarfs has yielded detections
of astrospheric absorption in six of nine cases.  The
67\% astrospheric detection percentage is consistent with the
high detection fraction previously found for stars within 7~pc
\citep{bew18}.  This in turn is consistent with the notion that
the local ISM within 7~pc is predominantly like the warm, partly
neutral ISM known to surround the Sun, as opposed to the
fully ionized ISM that more generally characterizes the LB.
It is therefore likely that the three nondetections of
astrospheric absorption are due to weak stellar winds, meaning
that we will be able to infer upper limits for the stellar
wind mass-loss rates for these nondetections.

     Estimating stellar mass-loss rates from astrospheric
absorption requires guidance from hydrodynamic models of
the astrosphere, analogous to models of the global heliosphere
that have been computed for decades to confront heliospheric
observations from spacecraft such as {\em Voyager} and
the {\em Interstellar Boundary Explorer} (IBEX).
Analogous to past studies \citep{bew05a,bew14}, the
code we use to model the astrospheres is a 2.5-D axisymmetric,
multi-fluid code that treats the plasma as a single fluid,
but the neutral hydrogen as multiple fluids corresponding to
distinct heliospheric regions in which charge exchange yields
different populations of neutrals, which do not equilibrate
with the plasma or with each other due to the low
densities \citep{gpz96}.  The neutral and plasma fluids
interact primarily through charge exchange interactions,
involving an electron jumping from a neutral H atom to a proton.

     The heliospheric model that has long represented the
starting point for our analysis is one that has consistently
demonstrated its ability to reproduce heliospheric absorption
detected for various lines of sight in various directions
while assuming plausible values for the ISM boundary conditions
of the heliosphere, including the local ISM's flow velocity
in the solar rest frame, $V_{ISM}=26$ km~s$^{-1}$
\citep{bew00,bew05b}.  We have already
noted its success in reproducing the heliospheric absorption
seen for three of our nine observed lines of sight.

     An astrospheric model assuming a solar mass-loss rate is
computed by changing $V_{ISM}$ in the heliospheric model to the
value appropriate for the star (see Table~3), while keeping
everything else the same.  In order to experiment with different assumed
stellar mass-loss rates, we simply vary the assumed stellar wind
density at the inner boundary, which is typically at 1~au from the
star.  From such models we can compute
the absorption predicted by the model by integrating along the
LOS through the astrosphere, with a direction defined by the
$\theta$ value in Table~3.  Higher $\dot{M}$ values naturally
lead to larger astrospheres, larger hydrogen wall column
densities, and therefore more Lyman-$\alpha$ absorption.
Rather than varying the stellar wind density to modify
$\dot{M}$, we could alternatively vary the stellar wind velocity, $V_w$.
However, along the lower main sequence stellar mass and radius happen
to vary in such a way that the surface escape speed is relatively
constant.  This provides some reason to believe that the winds of
lower main sequence stars might all have similar speeds, given that
the solar wind velocity is similar to the Sun's surface escape speed
of 618 km~s$^{-1}$.
The astrospheric absorption is to first order determined by the
size scale of the astrosphere, which depends on the stellar wind
ram pressure, $P_w$.  Since $P_w\propto \dot{M}V_w$, mass loss
rate estimates will vary inversely with the assumed $V_w$ \citep{bew02}.

     We have in the past estimated that $\dot{M}$ values
measured in this way should be accurate to within about a factor of two
\citep{bew05a}, with important systematic uncertainties
including the unknown degree of variation in ISM properties from
star to star, and possible differences in stellar wind speed.
The accuracy of the physics in the heliospheric/astrospheric
modeling code is another source of uncertainty, but this model
dependence is not as crucial as one might suppose, because the
approach of extrapolating the astrospheric models from a
heliospheric model that reproduces the heliospheric absorption
makes the procedure somewhat semi-empirical.  The physics of
how an astrosphere's size and Lyman-$\alpha$ absorption
properties respond to changes in $V_{ISM}$ and stellar wind
density is relatively simple, depending on the balance of ram
pressure between the two flows, so these changes should be
relatively insensitive to the details of the physics in the
model being used.  Thus, there is little reason to believe
that replacing our code with more sophisticated ones that
are now available would change our conclusions significantly.
More sophisticated models include ones that are fully 3-D,
including heliospheric and ISM magnetic fields, and ones
with a fully kinetic treatment of the neutrals
\citep[e.g.,][]{vvi09,nvp13,mo15}.

\begin{figure}[t]
\plotfiddle{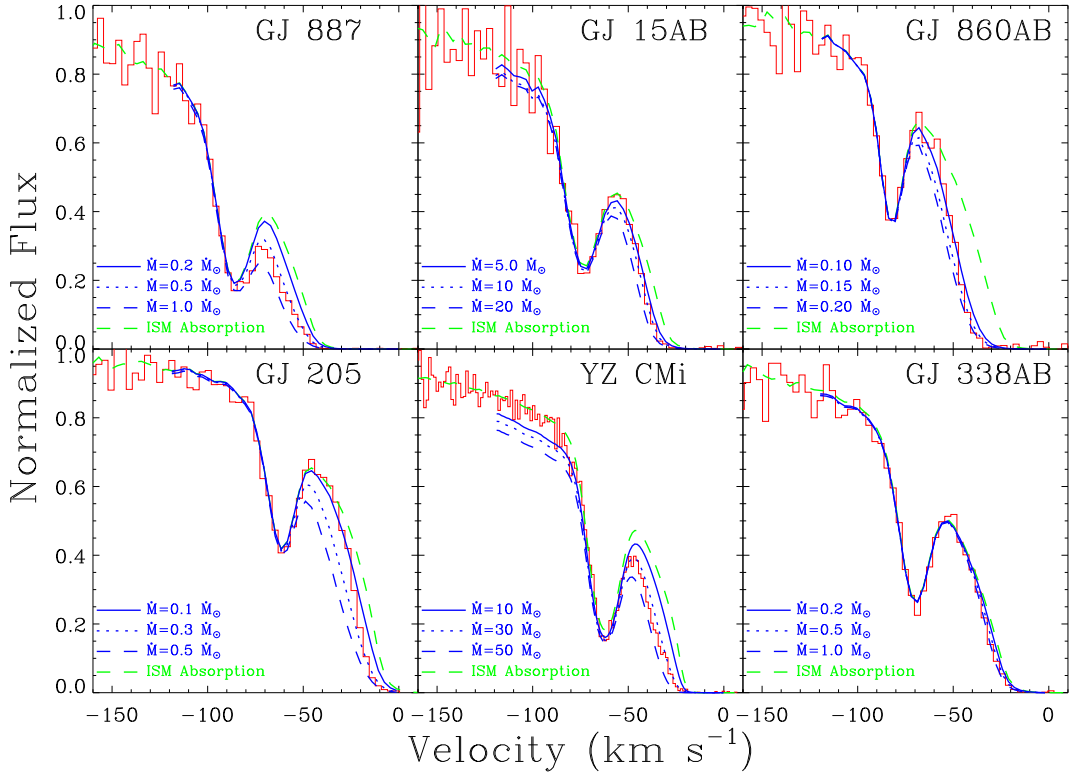}{3.5in}{0}{90}{90}{-290}{-355}
\caption{The blue side of the H~I Lyman-$\alpha$ lines for the
  six new astrospheric absorption detections.  The green dashed
  lines show the ISM absorption, and the blue lines show the
  additional astrospheric absorption predicted by models
  assuming various mass-loss rates.}
\end{figure}
     Figure~4 shows the Lyman-$\alpha$ lines of the six M dwarfs
with newly detected astrospheric absorption, zooming in on the
blue side of the H~I absorption profile where the astrospheric
absorption lies.  The figure also shows the absorption predicted
by a variety of astrospheric models with different assumed
mass-loss rates, after being added to the ISM absorption.
Astrospheric absorption naturally increases as $\dot{M}$
is increased, due to a larger astrosphere with a thicker
hydrogen wall and higher H~I column densities along the
observed LOS through the astrosphere.

     Judging which model best reproduces the observed absorption
involves some degree of subjectivity.  The discrepancy with the
data naturally varies along the side of the H~I absorption profile.
It is generally more important to fit the data near the base
of the absorption than higher along the side of the profile,
because reasonable adjustments to the shape of the assumed
stellar Lyman-$\alpha$ emission profile can in principle
improve discrepancies at higher flux levels.  Such corrections
are harder near the base of the absorption, as the necessary
adjustments to the stellar profile would introduce
implausible fine structure into the profile.

\begin{figure}[t]
\plotfiddle{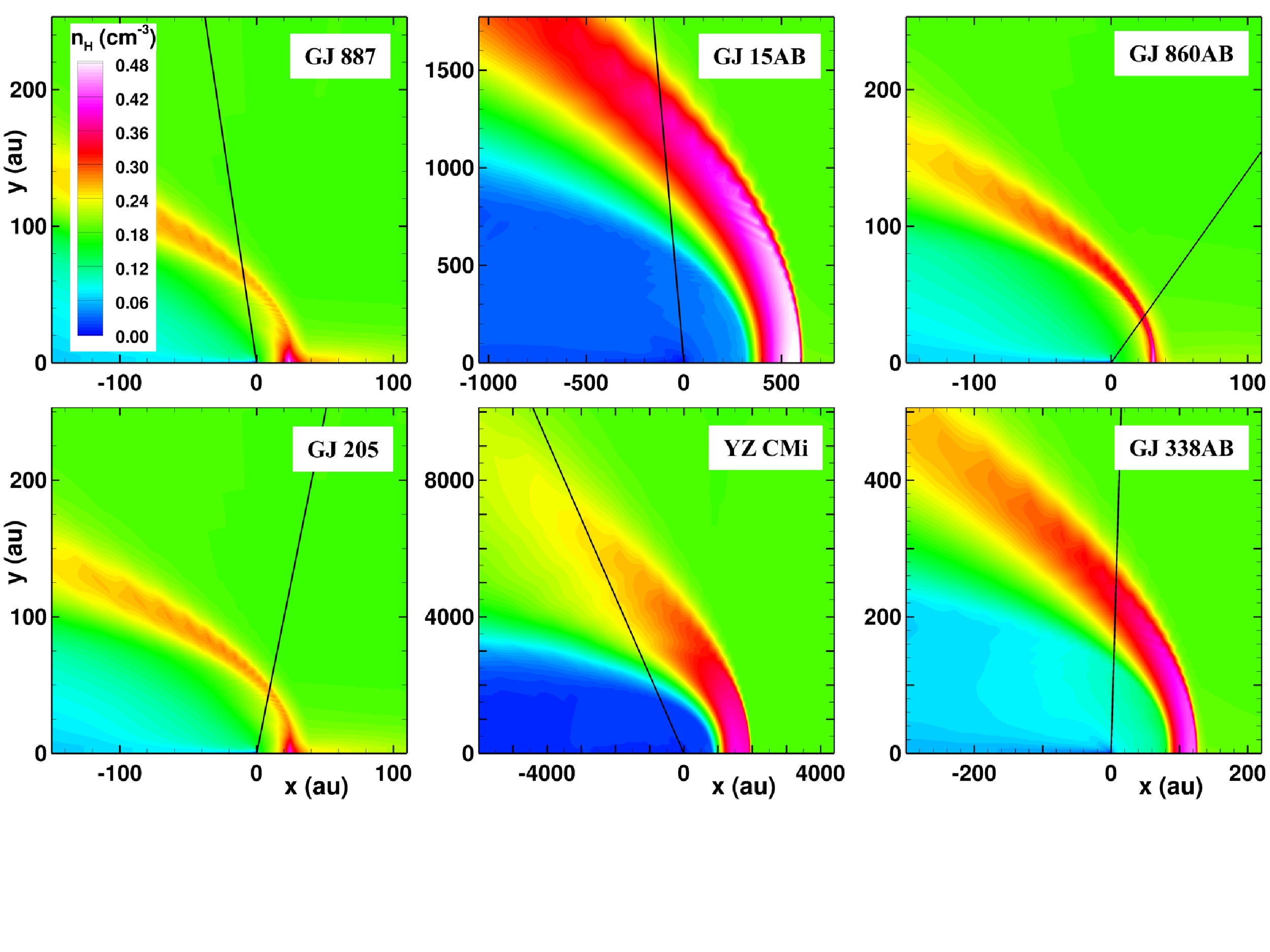}{3.5in}{0}{65}{65}{-230}{-70}
\caption{Maps of H~I density for the astrospheric models that
  yield the best fits to the data in Figure~4.  The laminar
  ISM wind seen by the star comes from the right.  Solid lines
  indicate the observed Sun-star line of sight.}
\end{figure}
     The $\dot{M}$ values that best fit the data are
listed in Table~3, and Figure~5 shows H~I density maps for
the best-fit astrospheric models.  The highest H~I densities
along the LOS to the star are in the hydrogen wall region,
which is the parabola-shaped reddish (or purplish-red) region
seen in each panel.  This is where the ISM material is piled
up outside of the stellar astropause, with the outer edge of the
hydrogen wall marking the location of the stellar bow shock.
It is this hydrogen wall region that generally dominates the
astrospheric absorption signature, not just due to the higher
H~I densities, but also due to the decelerated flow speeds and
relatively high temperatures
at that location.  As the ISM passes through the stellar bow
shock, it is heated as well as compressed and decelerated.
(It turns out that the situation is somewhat different for
YZ~CMi, as will be described in Section~4.3.)

     The size of the astrospheres in Figure~5 varies
tremendously, depending on both $V_{ISM}$ and $\dot{M}$.
The smallest astrosphere is that of GJ~205, with an
upwind bow shock distance of only about 30~au, due to
both a high ISM flow speed of $V_{ISM}=70$ km~s$^{-1}$
and a low mass-loss rate of $\dot{M}=0.3$ $\dot{M}_{\odot}$.
This can be compared with the huge astrosphere of YZ~CMi,
with an upwind bow shock distance of 2000~au, due to
both a low ISM flow speed of $V_{ISM}=20$ km~s$^{-1}$
and a high mass-loss rate of $\dot{M}=30$ $\dot{M}_{\odot}$.
It is from these best-fit astrosphere models that we can
confirm that the astrospheres of the binaries GJ~15AB,
GJ~860AB, and GJ~338AB are indeed large enough to encompass
both members of the binary, meaning that the stellar wind
that is being diagnosed is that of the combined winds of
both stars.

     Turning our attention to the astrospheric
nondetections within 7~pc, we note that there are eight of
these listed in Table~3.  For one of them, Proxima~Cen, there
is already a published upper limit of
$\dot{M}<0.2$ $\dot{M}_{\odot}$.  We now seek to infer
upper limits for the other seven stars.  For GJ~729 and
AD~Leo, we ultimately conclude that no useful $\dot{M}$
constraints can be inferred, primarily due to the very low
$V_{ISM}$ values for these stars (see Table~3).  This greatly
complicates the astrospheric modeling, in part due to the
kinds of physical effects discussed in the next subsection
about YZ~CMi, which also has low $V_{ISM}$ (though not
nearly as low as GJ~729 and AD~Leo).

\begin{figure}[t]
\plotfiddle{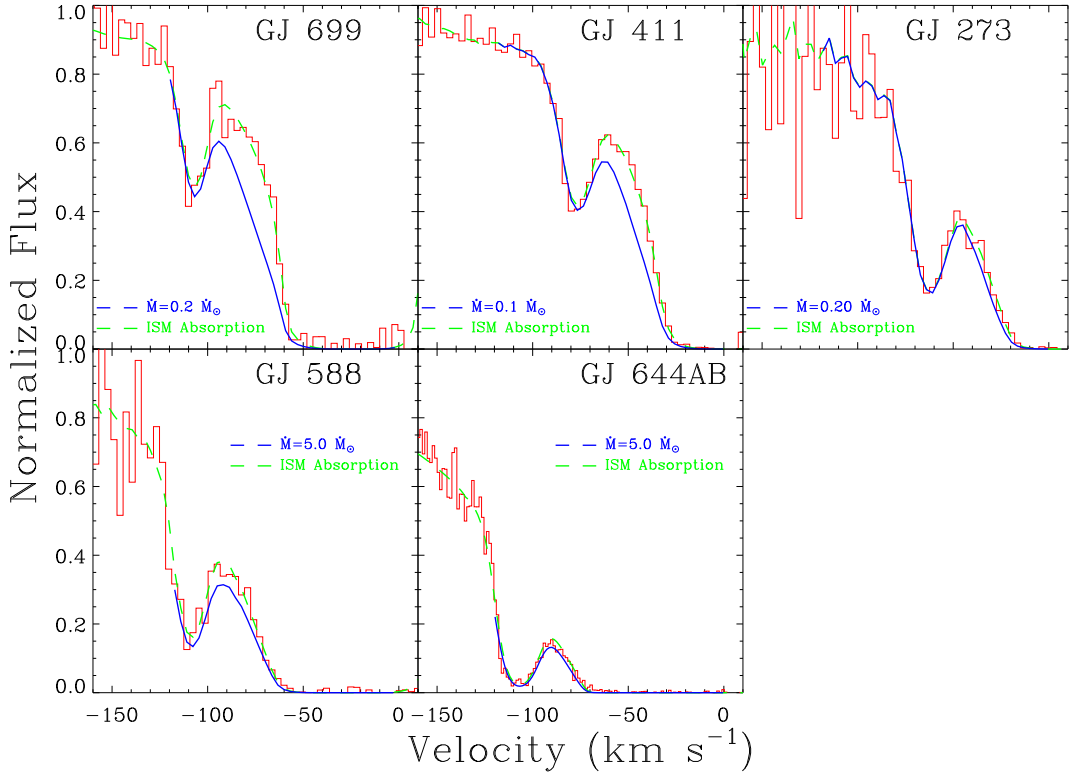}{3.5in}{0}{90}{90}{-290}{-355}
\caption{The blue side of the H~I Lyman-$\alpha$ lines for
  five astrospheric absorption nondetections, for which we
  infer upper limits for $\dot{M}$.  The green dashed
  lines show the ISM absorption, and the blue lines show the
  additional astrospheric absorption predicted by the models
  that produce just enough absorption that we deem the
  absorption to be likely detectable, thereby defining our
  $\dot{M}$ upper limit.}
\end{figure}
     The Lyman-$\alpha$ spectra of the remaining five
M dwarf astrospheric nondetections are shown in Figure~6,
where we are once again zooming in on the blue side of
the H~I absorption line where the astrospheric absorption
would be if any had been detected.  We note again that
the original analyses of GJ~699 and GJ~411 are presented
elsewhere (Youngblood et al., in preparation).  For each star,
models are constructed assuming different $\dot{M}$ to see how
large $\dot{M}$ must be for there to have been detectable
astrospheric absorption for the observed LOS.
Analogous to the situation for the astrospheric
detections, there is some subjectivity in deciding
how much excess astrospheric absorption must be predicted
beyond that from the ISM before it should be considered
detectable.  Figure~6 illustrates the predicted absorption of
the models that we decide represent the upper limits.
These range from $\dot{M}<0.1$ $\dot{M}_{\odot}$ for GJ~411 to
$\dot{M}<5$ $\dot{M}_{\odot}$ for GJ~588 and
GJ~644AB.

\begin{figure}[t]
\plotfiddle{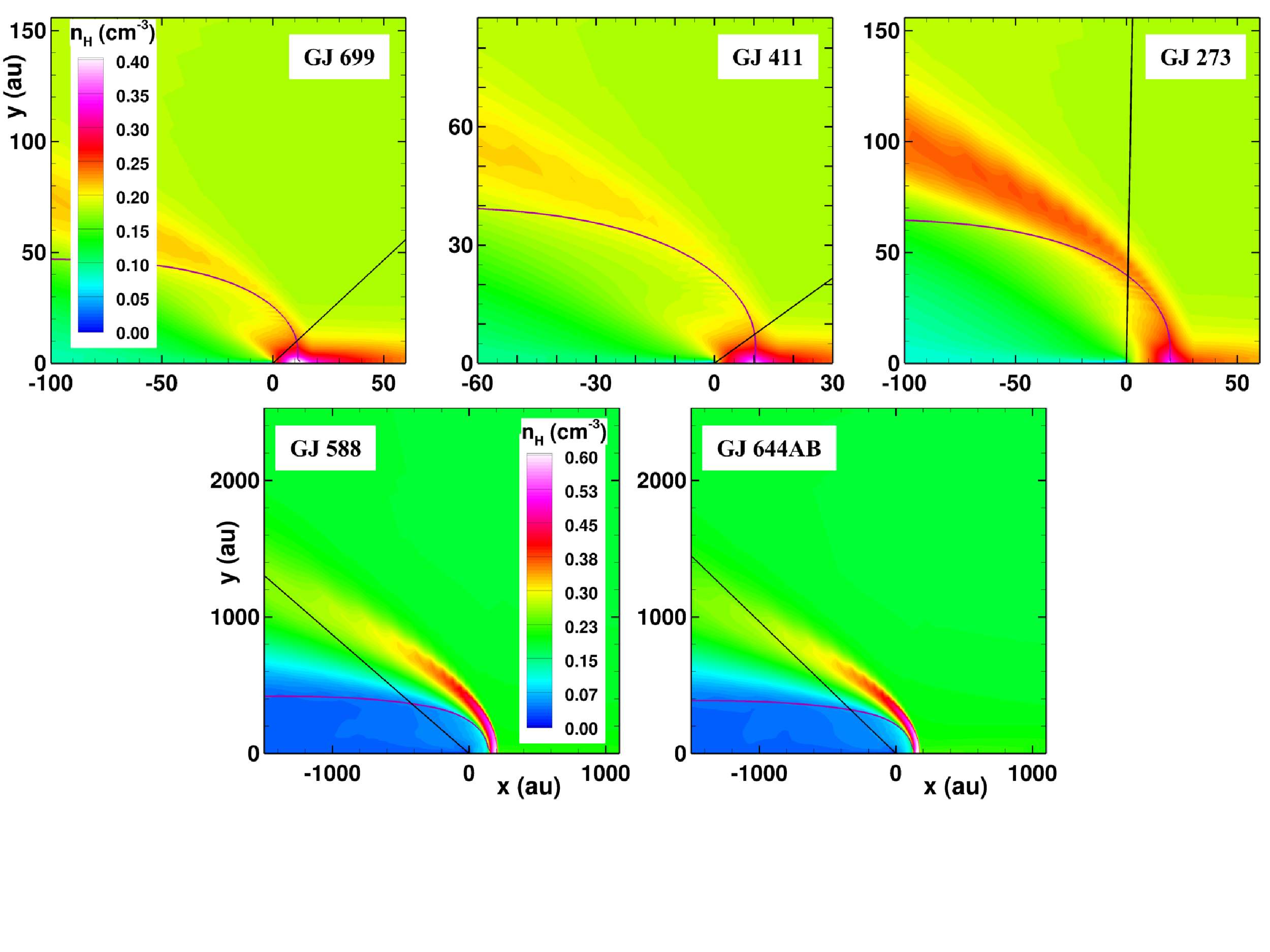}{3.5in}{0}{65}{65}{-230}{-70}
\caption{Maps of H~I number density for the astrospheric models that
  define the $\dot{M}$ upper limits in Figure~6.  The color bar in the
  GJ~699 panel applies to the three upper panels, and the color bar
  in the GJ~588 panel applies to the two lower panels.  Solid straight lines
  indicate the observed Sun-star line of sight.  The parabolic shaped
  astropause is also shown in each panel.}
\end{figure}
     The H~I number density maps of the upper limit astrospheric
models are shown in Figure~7.  The lines of sight to GJ~588 and GJ~644AB
are particularly downwind, which is not advantageous for detecting
astrospheric absorption, missing the most detectable part of
the hydrogen wall.  This in part explains the relatively high
upper limits for $\dot{M}$ for these stars.  The upper limit models for
GJ~699 and GJ~411 have upwind bow shock distances of only about 10~au,
due primarily to high $V_{ISM}$.  Considering that these
are upper limits, these astrospheres are clearly very small,
easily the most compact astrospheres inferred so far using
the astrospheric Lyman-$\alpha$ absorption diagnostic.

     Considering all the M dwarf $\dot{M}$ constraints from
Table~3 together, we now have actual $\dot{M}$ measurements
for nine stars, and upper limits for six.
For 13 of the 15 M dwarfs, the results are
consistent with the winds being comparable or weaker to
the solar wind.
Weak winds for M dwarfs are not surprising,
considering the small size of these stars.  However,
M dwarfs can be surprisingly active, with frequent and energetic
flares, so higher $\dot{M}$ values might have been
expected as well.  There are two M dwarf $\dot{M}$
measurements that clearly stick out for being unusually
high:  GJ~15AB with $\dot{M}=10$ $\dot{M}_{\odot}$, and
YZ~CMi with $\dot{M}=30$ $\dot{M}_{\odot}$.  Before discussing
the implications of these results further, it is necessary
to discuss some important and unique characteristics of
the YZ~CMi astrosphere.

\subsection{YZ~CMi: New Physics for Astrospheric Absorption}

     Heliospheric and astrospheric Lyman-$\alpha$ absorption has been
detected for many lines of sight toward nearby stars, despite H~I column
densities that are $3-5$ orders of magnitude lower than the ISM H~I
column densities for these lines of sight.  Astrospheric absorption is
nevertheless detectable because as the ISM approaches an astrosphere,
it is heated and decelerated as it nears the astropause, which is
the boundary separating the plasma flows of the stellar wind and the ISM.
The heating results in a broader absorption profile, which helps to
extend the astrospheric absorption in wavelength beyond that from the
ISM, and the deceleration shifts the absorption blueward, creating
the excess blue-side absorption that characterizes the astrospheric
absorption signature.

     Both the heating and deceleration effects are highly dependent on the
ISM flow speed seen by the star, $V_{ISM}$.  The Sun sees
$V_{ISM}=26.08\pm 0.21$ km~s$^{-1}$ \citep{bew15}.  This happens to
correspond to a flow with a Mach number of $M\approx 1$, leading to much
debate in the space physics community about whether there is a bow shock
in front of the heliosphere, or whether no real shock exists and that the
pile-up region outside the heliopause is better characterized as a
``bow wave'' \citep{gpz13}.  Even without a bow shock, there will
still be adiabatic deceleration, compression, and
heating of the material in the bow
wave, and there is no truly dramatic decrease in heliospheric absorption
for the bow wave case with $M\approx 1$ compared to a bow shock case with
$M\approx 1$ \citep{gpz13}.  Nevertheless, it is no accident that
nearly all the astrospheric detections listed in Table~3 have $V_{ISM}$
higher than the solar example.  This makes the existence of a bow shock far
less ambiguous than for the Sun, producing more heating and
deceleration of the ISM outside the astropause than the solar case,
making the astrospheric absorption more detectable.

     In selecting targets for our HST M dwarf observing program, we tried
to avoid stars with low $V_{ISM}$.  Nevertheless, an exception was made
for YZ~CMi.  For our sample, we wanted to include at least a couple very
active M dwarfs, which still had to be within 7~pc for reasons discussed
in Section~2.  The best available target after GJ~644 was YZ~CMi.
We have been rewarded
for this choice by the clear detection of astrospheric absorption for this
star (see Figure~3), despite the low $V_{ISM}=20$ km~s$^{-1}$ value.
Furthermore, our astrospheric models are indeed able to reproduce
the absorption well, with a sufficiently high mass-loss rate of
$\dot{M}=30$~$\dot{M}_{\odot}$ (see Figure~4).
However, examination of the
YZ~CMi astrospheric model reveals that the astrospheric absorption
has an origin somewhat different from other astrospheric detections,
which requires more discussion.

     Like all other cases of astrospheric absorption, hydrogen wall
neutrals are the source of the absorption, which are a population of
neutral H created by charge exchange with the ISM protons that have been
heated by passage through the bow shock (or by compression in the bow wave).
The difference for YZ~CMi is that the absorption is not coming primarily
from the hydrogen wall itself, where neutral H densities are highest (the
yellow and reddish region along the LOS in Figure~5), but from neutrals
very near and {\em inside} the astropause (the blue and green region along
the LOS in Figure~5).  The neutrals have an unexpectedly high temperature in
this location, explaining why they produce more absorption than the bulk of
the particle population in the hydrogen wall region.

\begin{figure}[t]
\plotfiddle{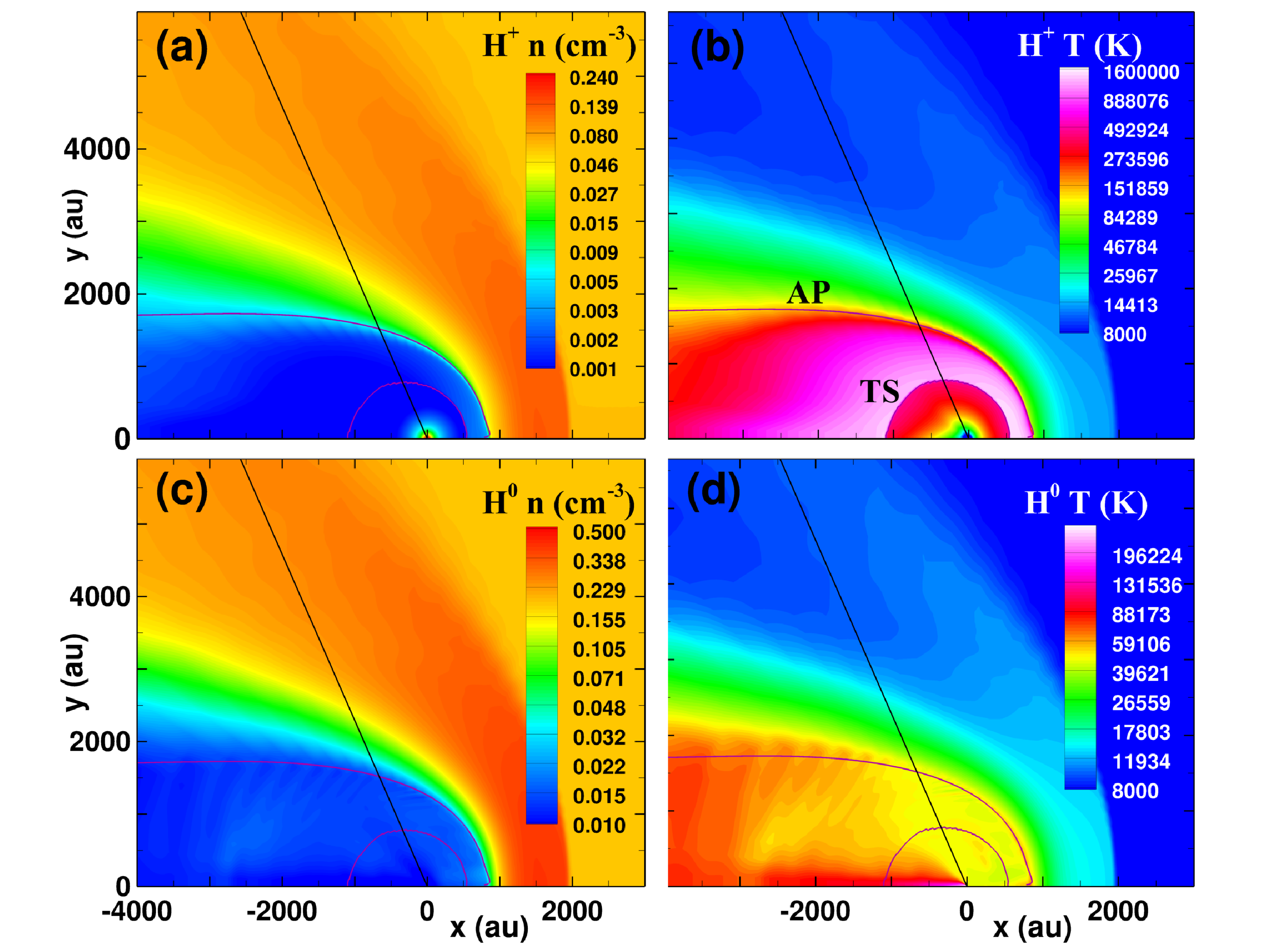}{3.5in}{0}{55}{55}{-200}{-15}
\caption{Maps of H$^+$ and H$^0$ number density and temperature for
  the $\dot{M}=30$~$\dot{M}_{\odot}$ YZ~CMi model.  Locations of the
  termination shock (TS) and astropause (AP) are indicated.  The straight
  solid black line indicates our LOS to the star.}
\end{figure}
     To explore this further, Figure~8 shows maps of number density and
temperature for protons (H$^+$) and for neutral H (H$^0$), for the
$\dot{M}=30$~$\dot{M}_{\odot}$ YZ~CMi model.  The H$^+$ and H$^0$ temperatures
within the hydrogen wall are generally rather low, $T\approx 15,000$~K.
Such temperatures are expected, given the low $V_{ISM}=20$ km~s$^{-1}$
speed of the ISM flow, and are not conducive to producing detectable
astrospheric absorption.  Another problem is the less-than-ideal crosswind
LOS to the star.  There will be some blueward shift of the hydrogen wall
flow relative to the ISM flow, due to deflection around the astrosphere,
but not as much as if the LOS were upwind.  In short, it is questionable
whether the YZ~CMi hydrogen wall could yield detectable astrospheric
absorption even for high values of $\dot{M}$.

     However, the $H^+$ temperature is actually surprisingly high just
outside the astropause, with $T\approx 6\times 10^4$~K.  Although densities are
significantly lower here, the large size of the astrosphere means that there
is still a significant amount of charge exchange happening, thereby creating
significant numbers of hydrogen wall neutrals with these high temperatures, some
of which can cross the astropause and dominate the part of the LOS inside
the astropause.  Since our multi-fluid code uses a single Maxwellian fluid
to represent all neutrals produced by charge exchange within the hydrogen
wall, the mixture of hot neutrals produced very close to the astropause with
the cooler neutrals produced in the bulk of the hydrogen wall yield H$^0$
temperatures along the LOS that are $T=30,000-50,000$~K near and inside the
astropause (see Figure~8d).  Despite the relatively low H$^0$ densities
in this area, the huge astrospheric size means that there is still
sufficient H~I column density for these neutrals to yield detectable
absorption.  This absorption exceeds that from the much denser, but much
cooler, hydrogen wall itself, and this is actually the source of the
absorption shown in Figure~4.

\begin{figure}[t]
\plotfiddle{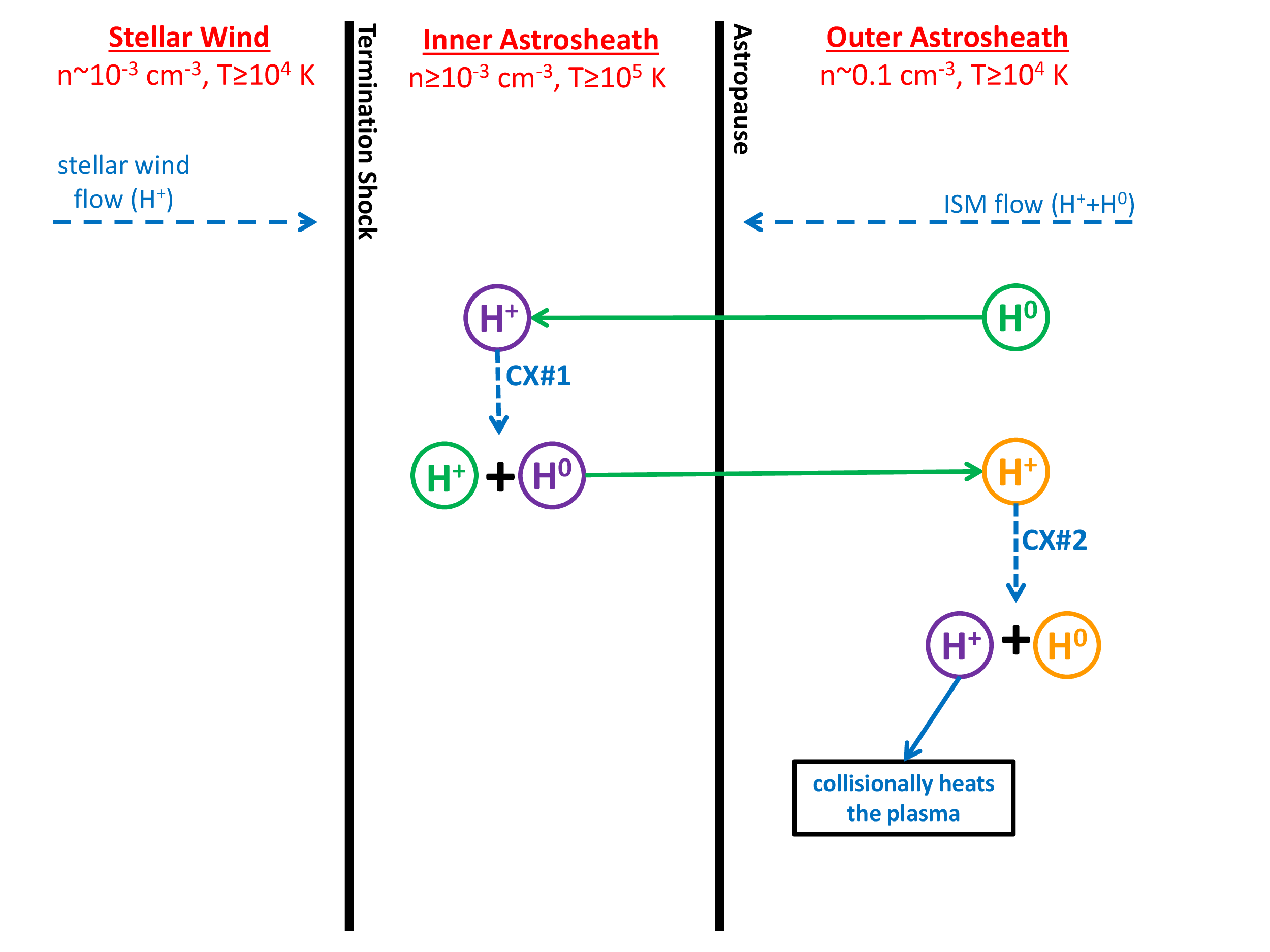}{3.5in}{0}{50}{50}{-160}{-10}
\caption{Schematic picture of YZ~CMi astrospheric structure, illustrating the
  sequence of two charge exchange interactions (CX\#1 and CX\#2) by which
  a hot proton in the inner astrosheath (in purple)
  can be transported across the astropause, and there heat the plasma.}
\end{figure}
     The conclusion is that the immediate source of the astrospheric
absorption signature for YZ~CMi is charge exchange with
this surprisingly hot H$^+$ just outside
the astropause.  But what is heating the H$^+$ there?  The answer is that
it is being heated by outward heat transport across the astropause from
neutrals created by charge exchange {\em inside} the astropause, some of
which then cross the astropause and dump their energy outside of it via
another charge exchange interaction.  Figure~9 provides an illustration of
the sequence of two charge exchange interactions that allow this to
happen.  The inner astrosheath region just
inside the astropause is very hot, as it is characterized by stellar wind
that is heated significantly by its passage through the termination shock.
However, this region has very low densities (see Figure~8a).  Although ISM
neutrals can penetrate the astropause and charge exchange there, in most
cases there are too few of these interactions for them to be important.
However, the YZ~CMi astrosphere in Figure~8 is so large that there are
enough of these inner astrosheath neutrals created to transport significant
amounts of energy out of the inner astrosheath and into the part of the
hydrogen wall just outside the astropause, with the energy deposited via
another charge exchange reaction.  Potential effects of this kind
of anomalous heating have been discussed before in the context of
heliospheric models \citep[e.g.,][]{gpz13}, but with YZ~CMi we have for the
first time encountered a case where this heating across the astropause has
become more important than any heating occurring within the actual
bow shock/wave, with regards to yielding detectable H~I absorption.

     One downside to the different physics involved in the YZ~CMi astrospheric
absorption signature is that it undoubtedly increases the uncertainty in
our $\dot{M}$ measurement.  The arguments in Section~4.2 about the
relative unimportance of model dependence for the $\dot{M}$ measurements
no longer apply, since the physics of the YZ~CMi absorption is
significantly different from the physics of the heliospheric Lyman-$\alpha$
absorption in the baseline heliospheric model, from which all of the
astrospheric models are basically extrapolated.  This is definitely
a case where further modeling efforts would be worthwhile, particularly
models that include a fully kinetic treatment of the neutrals.  The
enhanced uncertainty in the YZ~CMi measurement is problematic
considering how important this measurement potentially is, representing
the highest $\dot{M}$ per unit surface area yet detected via the
astrospheric Lyman-$\alpha$ absorption diagnostic.  Nevertheless,
in our discussion of the ramifications of our new $\dot{M}$ measurements
in the next section, the high YZ~CMi $\dot{M}$ measurement will be
assumed valid.

\section{Implications of the M Dwarf Wind Measurements}

\subsection{Relating $\dot{M}$ and Coronal Activity}

\begin{figure}[t]
\plotfiddle{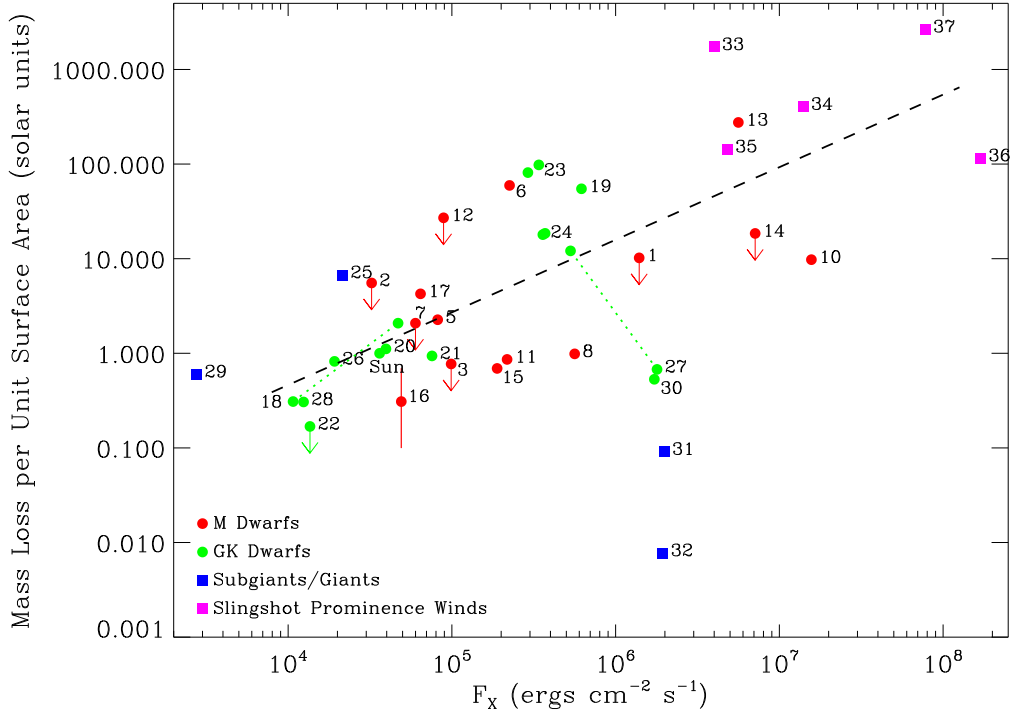}{3.5in}{0}{80}{80}{-260}{-310}
\caption{Mass-loss rate per unit surface area plotted versus X-ray surface
  flux for coronal winds, with the stars identified by the ID numbers from
  Table~3.  Dotted lines connect members of binary systems, with the assumed
  distribution of wind from Table~3.  Most of the constraints are from the
  astrospheric Lyman-$\alpha$ absorption diagnostic, but we have added
  slingshot prominence wind measurements from
  Jardine \& Collier Cameron (2019).  The M dwarf with
  the error bar is an $\dot{M}$ constraint from Lyman-$\alpha$ absorption
  seen during an exoplanet transit for GJ~436 (Vidotto \& Bourrier 2017).
  A power law of $\dot{M}\propto F_X^{0.77\pm 0.04}$ is fitted to the data
  points, excluding the subgiant/giant stars.}
\end{figure}
     With the new M dwarf $\dot{M}$ measurements provided here, we can
try to characterize the winds of M dwarfs for the first time.  This is
most naturally done in a plot like Figure~10, showing mass loss rate
per unit surface area plotted versus X-ray surface flux.  This is done not
only for the M dwarfs, but for the other stars listed in Table~3 as well.
For the solar data point, the mean solar X-ray luminosity assumed is from
\citet{pgj03}.  This would vary by about a factor of 4
over the course of the activity cycle \citep{tra20}.

     Trying to relate $\dot{M}$ with $F_X$ is natural, given that hot
coronae represent the source regions of coronal winds.  The existence
and basic characteristics of the solar wind can to first order be
understood as simple thermal expansion from the hot solar corona
\citep{enp58}.  Nevertheless, it is far from obvious a~priori that
$\dot{M}$ and $F_X$ should be correlated, given that coronal X-ray
emission originates from coronal loops, which are closed magnetic
fields with both footpoints tied to the photosphere, while coronal
wind will be coming from open field regions with only one footpoint
tied to the photosphere and the other end of the field line extending
out into the astrosphere.  For the Sun, no correlation
between $\dot{M}$ and $F_X$ is observed during the solar activity
cycle \citep{oc11}.

     Nevertheless, Figure~10 seems to show a general trend of increasing
$\dot{M}$ with $F_X$, albeit with lots of scatter.  A power law fit to the
data points, excluding the subgiant/giant stars, yields
$\dot{M}\propto F_X^{0.77\pm 0.04}$, as shown in the figure.  This
is a somewhat flatter relation than the $\dot{M}\propto F_X^{1.34\pm 0.18}$
result previously reported for the GK stars alone, excluding $\xi$~Boo~A and
$\pi^1$~UMa \citep{bew05a}.  The impression of
a wide range of $\dot{M}$ at a given $F_X$ value is greatly increased
with the inclusion of the numerous new M dwarf data points.
The M dwarf measurements suggest that main sequence stars with a
given $F_X$ value can have $\dot{M}$ values that vary by up to two
orders of magnitude.  This impression relies heavily on the two M
dwarfs with surprisingly strong winds, GJ~15AB (star \#6) with
$\dot{M}=10$~$\dot{M}_{\odot}$ and YZ~CMi (star \#13) with
$\dot{M}=30$~$\dot{M}_{\odot}$.  The GJ~15AB measurement is well over
an order of magnitude higher than the $\dot{M}=0.15-0.5$~$\dot{M}_{\odot}$ 
measurements for GJ~860AB, GJ~205, and GJ~338AB (stars \#8, \#11, and \#15),
which have similar $F_X$.  At a higher activity level, the YZ~CMi
measurement is 30 times higher than that of EV~Lac (star \#10),
despite similar $F_X$.

     The YZ~CMi data point is also notable for being the one astrospheric
Lyman-$\alpha$ measurement that overlaps with the region of Figure~10
occupied by the slingshot prominence winds.  The slingshot prominence
stars are all characterized by extremely rapid rotation.  The fastest
rotator among the stars with astrospheric measurements is indeed YZ~CMi, with
$P_{rot}=2.78$ days \citep{eda19}, but this is still much
slower than the slingshot prominence stars, with $P_{rot}\leq 0.5$ days
\citep{mj19}.  Nevertheless, \citet{cvd18}
find that both YZ~CMi and EV~Lac are
rotating fast enough to be in the centrifugal confinement regime that could
allow the slingshot prominence phenomenon to exist, and they estimate
potential mass loss rates for YZ~CMi and EV~Lac of up to
$\dot{M}=0.7$ $\dot{M}_{\odot}$ and $\dot{M}=0.28$ $\dot{M}_{\odot}$,
respectively.  These are lower than our measurements, and therefore
consistent.  It is possible that some fraction of the wind that
we are detecting for YZ~CMi and EV~Lac could consist of slingshot
prominence material.

     Considering only the GK dwarfs in Figure~10, the impression is of a
comparatively tight relation between $\dot{M}$ and $F_X$ for $\log F_X<6.0$,
with the two surprisingly low $\dot{M}$ values with $\log F_X>6.0$ suggesting
a change of behavior in the $\dot{M}$-$F_X$ relation for $\log F_X>6.0$.
In the past, this had been referred to as a possible ``wind dividing line''
\citep{bew05a,bew14}.  The existence of such a dividing line
is clearly weakened by the new M dwarf measurements.  With their inclusion,
the two low GK dwarf data points no longer appear clearly inconsistent
with the high GK dwarf measurements seen at slightly lower activity, given
the high degree of scatter apparent in the loose $\dot{M}$-$F_X$ relation.

     The $\dot{M}$ measurements available to date now imply
that coronal wind strength is not dependent solely on spectral type and
coronal activity, though this inference relies heavily on just two
of the new M dwarf measurements.
Without the GJ~15AB and YZ~CMi data points, Figure~10 would suggest
that M dwarfs all have rather low $\dot{M}$, regardless of activity,
and their winds may be exhibiting different behavior compared to the
GK dwarf winds at moderate to high activity.  With only two
strong M dwarf winds found to date, it would be helpful in the future
if other examples of strong M dwarf winds could be found, to provide
support for the GJ~15AB and YZ~CMi measurements.

     Our wind measurements can be compared with predictions from
theory.  For example, \citet{src11} use an Alfv\'{e}n wave
driven wind model to predict mass-loss rates for a wide range of cool
stars.  However, their predictions for M dwarfs are
very low ($\dot{M}<0.005$ $\dot{M}_{\odot}$), clearly inconsistent with
our results.  \citet{src11} suggest that M dwarf winds
may actually be dominated not by Alfv\'{e}n wave driving, but by
CMEs, a possibility we will return to in
Section~5.4.  Other Alfv\'{e}n wave based models infer more substantial
M dwarf winds, without resorting to CMEs \citep{aav14,jdag16,alm20}.
These models predict increases in wind
strength with stellar activity qualitatively consistent with the
observed relation in Figure~10, although quantitative comparison with
the models is complicated by the extensive scatter in the data.
For example, \citet{tks13b} predict $\dot{M}\propto F_X^{0.82}$,
in good agreement with the observed $\dot{M}\propto F_X^{0.77\pm 0.04}$.
It is possible that mass loss might depend on magnetic complexity as
well as disk-averaged magnetic flux \citep{cg15},
which could in principle contribute to the scatter.

\subsection{The Effects of Wind Variability}

     One possible interpretation of the scatter in the relation
between coronal activity and wind strength seen in Figure~10 is that
this is due to wind variability.  In this interpretation, the
discrepancy between the winds of the similar stars EV~Lac and YZ~CMi
is simply due to temporal variability, rather than being indicative
of any fundamental difference in wind behavior for these stars.  Evalulating
this possibility requires an evaluation of the timescale over which our
wind measurements are applicable.

     This timescale will depend on
the length of time it takes a wind ram pressure signal to reach the
hydrogen wall region well beyond the astropause.  This naturally
depends on the size of the astrosphere.  For the solar example,
it takes the solar wind roughly 6 months to make it to the termination
shock, at which point it is decelerated to even slower speeds.  This
means that it takes many years for any change in solar wind
pressure to register at the heliopause.  Time-dependent modeling
of the heliosphere \citep[e.g.,][]{nvp13} suggests only a very
weak variation in heliopause distance even on activity cycle timescales,
and heliospheric absorption is coming from even further distances from
the Sun.  As a consequence, we do not expect there to be any observable
change in heliospheric Lyman-$\alpha$ absorption even on decadal
timescales, let alone shorter ones.

     Returning to the EV~Lac/YZ~CMi comparison, the EV~Lac astrosphere is
somewhat more compact than the heliosphere
\citep{bew05a}, due to higher $V_{ISM}$, but the YZ~CMi astrosphere
is much bigger (see Figure~8).  Our large $\dot{M}$ value for YZ~CMi will be
characteristic of the average mass loss over a period of many decades, if
not centuries, depending on the exact stellar wind speed.  If the
astrosphere of YZ~CMi has ever been smaller and characteristic of a
weaker wind like that of EV~Lac, it has not been for a long time.
In general, attributing the scatter in Figure~10 to wind variability
seems unlikely due to the lack of sensitivity of the astrospheric wind
measurements to short timescale variability.


\subsection{Relating $\dot{M}$ and Magnetic Topology}


     If spectral type and coronal activity level are not solely
determinative of $\dot{M}$, what is the missing factor?  One likely
candidate is coronal topology.  It is possible to envision two
stars with similar $F_X$ values, but with the coronal emission
and associated magnetic field distributed very differently
across the stellar surface, which could lead to different
wind properties.  Exploring this possibility requires
knowledge of the magnetic field topology of our sample of stars.
Fortunately, information about this is in fact available for
many of the stars in the Table~3 sample, thanks to spectropolarimetric
measurements \citep{jm08b}.  A recent survey of such observations for
M dwarfs is provided by \citet{ok21}.  In this section, we try to relate
our wind measurements to field properties inferred by the
spectropolarimetric analyses.

     The M dwarfs in our sample with spectropolarimetric constraints
are EV~Lac \citep{jm08b}, YZ~CMi \citep{jm08b}, Proxima~Cen \citep{bk21},
and GJ~205 \citep{emh16}.  There are nine other stars in Table~3 with
spectropolarimetric measurements.  These include studies of
61~Cyg~A, $\epsilon$~Eri, $\epsilon$~Ind, $\pi^1$~UMa, and $\xi$~Boo~AB,
which \citet{aav16b} previously sought to relate to wind properties.
And finally, we also consider studies of GJ~892
\citep{cpf18}, $\lambda$~And \citep{dof21}, and
V347~Peg \citep{jm08a}, this last star being one of the slingshot
prominence stars.

\begin{figure}[t]
\plotfiddle{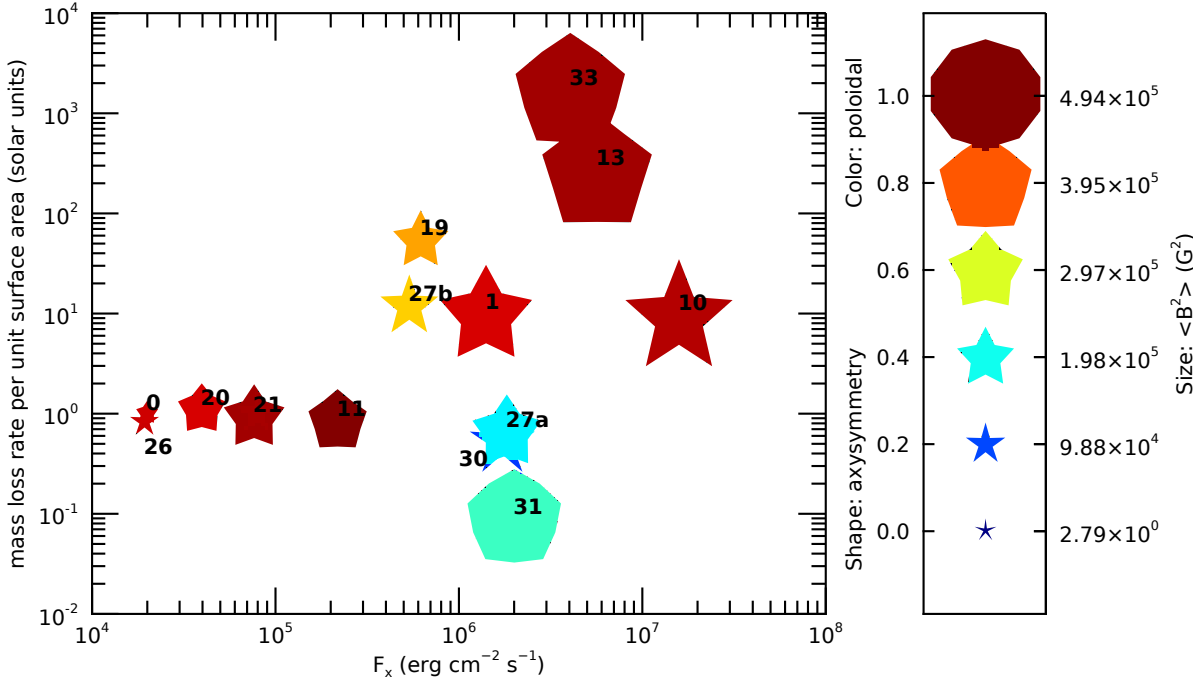}{3.2in}{0}{80}{80}{-250}{-190}
\caption{A reproduction of Figure~10, but symbols are used to indicate
  properties of the magnetic topology inferred from spectropolarimetry,
  where available.  As indicated by the key to the right of the plot,
  the size of the symbol indicates the magnitude of the square of the
  surface field, the color of the symbol indicates the relative amount
  of poloidal and toroidal field (with dark blue being purely toroidal
  and dark red being entirely poloidal), and the shape of the symbol
  indicates the degree of axisymmetry (with a decagon representing a
  purely axisymmetric field and a star with narrow arms representing a
  symmetric field).  A point is plotted for a solar minimum Sun (star \#0).}
\end{figure}
     Figure~11 provides an illustration of the spectropolarimetric
differences amongst the stars in our sample.  It is essentially a reproduction
of Figure~10, but with symbols used to indicate various topological properties
of the stellar magnetic field inferred from the spectropolarimetry.  This
is an updated version of a figure from \citet{aav16b}.  As described
by \citet{jd09}, the size of the symbols indicates the
logarithm of the average squared magnetic field ($\langle B^2 \rangle$),
the color of the symbols indicates the relative importance of poloidal
and toroidal fields, and the shape of the symbols indicates the degree
of axisymmetry of the field.  A point is plotted for the Sun at solar minimum
\citep{aav16a}.

     As expected, the more active stars with the higher $F_X$ values tend
to have the stronger fields, as indicated by the larger symbol sizes in
Figure~11.  There is nevertheless a very wide range in $\dot{M}$ among these
more active stars.  The three stars with the weakest winds
(stars \#27a, \#30, and \#31) have the most toroidal dominant
fields, suggesting a possible connection between toroidal fields and
weak winds.  As for the M dwarf stars that are our focal
point here, the EV~Lac/YZ~CMi comparison is once again of particular
interest (stars \#10 and \#13).  The biggest spectropolarimetric 
difference between these two stars lies in the inferred degree of
field axisymmetry, with YZ~CMi possessing a very axisymmetric field, and
EV~Lac exhibiting a strong departure from axisymmetry.  It is EV~Lac that
seems more unusual in the \citet{jm08b} sample of M dwarfs,
implying that YZ~CMi may be more typical of very active M dwarfs.
However, EV~Lac also is a somewhat slower rotator than the other stars
in the sample, with $P_{rot}=4.37$~days.
\citet{vs19} estimate similar magnetic filling factors for
EV~Lac and YZ~CMi, although their analysis does not distinguish between open
and closed magnetic fields.

     The picture of the YZ~CMi field provided by \citet{jm08b} is
that of a highly dipolar, axisymmetric field.
However, the spectropolarimetric data is open to interpretation, and
\citet{ds14} provide a somewhat different picture of the
overall field topology of YZ~CMi.  In their analysis, they find evidence
for a significant zero-field component for the surface of YZ~CMi, in
contrast to the other active M dwarfs in their study.  This could be
interpreted as a stellar analog for coronal holes, which on the Sun are
locations of weak surface fields and open field lines where high speed
wind streams are escaping.

     The EV~Lac/YZ~CMi dichotomy is not the only example of two seemingly
similar active M dwarfs with surprisingly different field properties.
Another example is the GJ~65AB binary (BL~Ceti+UV~Ceti), consisting of two
very similar, very active mid-M dwarfs, which nevertheless have very different
magnetic topologies \citep{ok18}.  This difference is likely
connected to differences in X-ray and radio behavior for these two coeval stars,
which has persisted for decades \citep{deg81,ma03}.  Ideally,
wind measurements would be made for GJ~65A and GJ~65B separately to see how
the differences in these coeval stars impact their winds, but the astrospheric
absorption technique will not work,
as the two stars are close enough to share the same astrosphere, meaning
that only the combined strength of the two stellar winds could be measured. 

     Besides EV~Lac and YZ~CMi, the only other two M dwarfs in our
sample with both astrospheric $\dot{M}$ constraints and spectropolarimetric
measurements are GJ~205 (star \#11) and Proxima~Cen (star \#1),
which are less active and have much longer rotation
periods.  The former has a very poloidal, axisymmetric field
topology, analogous to YZ~CMi, albeit with a much weaker overall field
strength, as expected for a more inactive star with a lengthy
rotation period of $P_{rot}=33.6$~days.  Despite the topological
similarities with YZ~CMi, the stellar wind of GJ~205
($\dot{M}=0.3$ $\dot{M}_{\odot}$) is over two orders of magnitude weaker
than that of YZ~CMi per unit surface area.  Like GJ~205, Proxima~Cen also has a
long rotation period ($P_{rot}=89.8\pm 4.0$ day) and a wind much weaker than that
of YZ~CMi ($\dot{M}<0.2$ $\dot{M}_{\odot}$), but unlike GJ~205 its magnetic
topology more resembles EV~Lac, with significant deviation from axisymmetry.

     The large scale field topology of a star is ultimately tied to
the properties of its internal dynamo, which also determines the nature
of any long-term activity cycle it might have.  In studies relating
activity cycle periods with rotation rates, there is a suggestion
of a bimodality in dynamo operation for stars with relatively fast
rotation periods of $P_{rot}<22$~days.  The two modes consist
of an inactive branch with very short activity
cycles ($P_{cyc}\leq 5$~yr) and an active branch
with much longer cycles \citep{ab98,tsm16}.
There is a hint of possible bimodality in Figure~10 as well for the
more active stars, separating ones with low $\dot{M}$ and those with
high $\dot{M}$.  This impression is particuarly acute for the M dwarfs,
with GJ~15AB and YZ~CMi having winds over an order of magnitude stronger
than all the other M stars.  It is worth noting that EV~Lac has a
reported short activity cycle of
$P_{cyc}\approx 5$~yr \citep{lnm86}, while the similarly
active YZ~CMi, with a much stronger wind,
has a much longer activity cycle of
$P_{cyc}=27.5$~yr \citep{nib18}.

     Clearly, more work is needed to explore
how $\dot{M}$ and magnetic topology might be
related for stars of various spectral types and activity levels.
Particularly valuable would be spectropolarimetric measurements
for GJ~15AB to try to explain why that binary has such a strong wind.

\subsection{Are M Dwarf Winds Dominated by CMEs?}

     The solar wind is characterized by a more or less continuous, steady
flow of plasma from the corona.  However, there is a transient component
to the wind, namely the CMEs.  The mass lost from the Sun due to CMEs is
over an order of magnitude less than the Sun's total mass loss
\citep{wm19}, but the CMEs
are still numerous enough to be important in many contexts, such as
space weather impacts on the Earth and other planets, particularly at times
of solar maximum.  Solar flares are often associated with CMEs, though it is
important to note that there are slow CMEs that occur with no associated
flare (and indeed no associated surface activity at all), particularly
at solar minimum \citep{bew17}.  Conversely, although it is true
that strong M and X-class flares on the Sun are usually found to be
associated with a fast CME emanating from above the flare site, this
is not always the case \citep{xs15}.

     For the Sun, a strong correlation is found between flare strength
as quantified by X-ray luminosity and CME mass \citep[e.g.,][]{ana11}.
Given that we have very
limited observational knowledge about the nature of CMEs emanating
from other stars, it is natural to apply solar flare/CME
relations to active stars that flare more frequently and energetically,
in order to estimate what CMEs might contribute to the stellar winds of
these stars \citep{spm19}.  However, doing this for truly
active stars invariably leads to conclusions that such stars should
have winds hundreds or thousands of times stronger than the solar
wind simply due to CMEs alone \citep[e.g.,][]{jjd13,po17}.  Such
conclusions are in obvious conflict with our astrospheric wind
measurements, which suggest only modest mass-loss rates for M dwarfs.

     Young, rapidly rotating, active M dwarfs are well known for
particularly frequent and energetic flaring, and EV~Lac and YZ~CMi
within our sample of M dwarfs are two of the most notorious sources
of massive flares.  These include true superflares
\citep[e.g.,][]{afk10}, including one for EV~Lac that triggered
a gamma ray burst detector, and is estimated to have produced X-ray
fluxes at flare peak that exceeded the quiescent bolometric luminosity
of the star \citep{rao10}.  We note that a large
UV flare occurred right at the end of our STIS/E140H YZ~CMi
observation, in addition to a few smaller flares.  Given the frequency
of such activity, as demonstrated further in the recent study of
\citet{hm21}, even the relatively large
$\dot{M}=30$ $\dot{M}_{\odot}$ value that we find for YZ~CMi seems
surprisingly modest, let alone the older $\dot{M}=1$ $\dot{M}_{\odot}$
measurement for EV~Lac.

     Clearly, the strong connection between flares and fast, massive
CMEs on the Sun cannot extend to flare stars like EV~Lac and YZ~CMi.
For such stars, CMEs must be far less common or far less massive
than one might expect,
given the frequent flaring.  There are other observations that support
this conclusion.  Attempts to detect radio Type~II bursts from
flaring M dwarfs, associated with CME shocks, have so far proved
unsuccessful \citep{mkc18a,mkc18b,jv19}.
Another observational signature
that has been occasionally observed during flares and interpreted as
indicating a CME eruption is Doppler shifted emission or absorption
in optical hydrogen Balmer lines during flares.  As a CME signature this
is imperfect, as it is more of a signature of an erupting prominence
than a CME, but on the Sun such prominence eruptions often end up
embedded within CMEs \citep{bew17}.
Systematic attempts to detect Balmer signatures have generally found
that they are very rare, and this rarity could indicate that stellar
CMEs are not as frequent from active stars as one might think
\citep{ml20,pm20,po20}.

     The Sun itself provides examples of what may be happening on
active flare stars, as there are many solar cases of strong flares
with no associated CMEs at all.  The most recent and best studied
examples are flares from active region AR~12192.  This was the biggest
active region of the last solar cycle, and the most productive of
strong flares, particularly in 2014~October.  But almost none of
the flares from AR~12192 had associated CMEs \citep{xs15,jkt15}.
On the Sun this behavior is unusual, but on active stars perhaps this
is the norm.  The cause of this may be strong magnetic field overlying
active regions that confines the flare and inhibits CME eruption.
Numerical simulations of CMEs on active stars made in recent years
include models of such confined eruptions \citep{jdag18,jdag19a,jdag20b}.

     Despite the difficulties in detecting CMEs on other stars,
we still cannot rule out the possibility that the stellar
winds that we are detecting for M dwarfs are CME-dominated.
Our measurements will typically be indicative of the average wind ram
pressure over years, if not decades (see Section~5.2), and there is
currently no way to tell whether this ram pressure signal is
dominated by CMEs or quiescent wind.

\subsection{Implications for M Dwarf Exoplanet Habitability}

     One important application of our new M dwarf wind measurements
is to better understand the environment of exoplanets around such
stars.  Remarkably, nearly half of the M dwarfs listed in Table~3 are
already known exoplanet hosts (Proxima~Cen, GJ~699, GJ~411, GJ~887,
GJ~15A, GJ~273, GJ~338B, and GJ~436).  Planets in habitable zones around
M dwarfs are of particular interest.  Due to the ubiquity of M dwarfs,
it seems likely that most planets in stellar habitable zones within the
Galaxy will  be orbiting M stars \citep{cdd15}.  One
of these in fact exists around our nearest stellar neighbor, Proxima~Cen
\citep{gae16}.  

     The habitable zones of M dwarfs are much closer to the stars
than for later type stars, so planets in such locations will potentially
be exposed to much higher particle fluxes from stellar winds.
Assessments of the potential impact of this wind exposure on planets in
M dwarf systems have been underway for some
time \citep[e.g.,][]{aav13,cg16,cd17,jdag19b,jdag20a}, and
these studies will greatly benefit from the constraints on M dwarf
winds provided here.

     The question of whether M dwarf habitable zone planets are truly
habitable is in part tied to the question of whether intense exposure
of such stars to stellar flares, CMEs, and energetic particles would
make habitability impossible \citep[e.g.,][]{mlk07,ay17}.
However, we have noted in the previous
section that CMEs from M dwarfs may be much less common than generally
thought, despite the high flare rate, so perhaps CME exposure is not
as big a factor for habitability as often supposed.  Furthermore, since
in the solar example damaging interplanetary energetic particles
originate from CME shocks rather than from flares, if fast CMEs are
less common than generally thought, perhaps energetic particle fluxes
are also lower \citep{ff19}.
Exoplanets in M dwarf habitable zones will certainly
be exposed to high X-ray fluxes, both from quiescent
coronal emission and flares.  There is no avoiding that, but it
remains highly questionable whether CMEs and energetic particles from
stellar activity are a major factor.




\section{Summary}

     We report on the results of an analysis of H~I Lyman-$\alpha$ lines of
nine M dwarf stars observed by HST/STIS, for purposes of studying
astrospheric Lyman-$\alpha$ absorption and estimating $\dot{M}$ for
these stars.  Our results allow us to truly characterize the coronal
winds of M dwarfs for the first time.  Our findings are summarized as
follows:
\begin{description}
\item[1.] Six of our nine HST/STIS targets yield successful detections
  of astrospheric Lyman-$\alpha$ absorption.
  This high detection fraction is consistent with previous
  studies of stars within 7~pc, consistent with the idea that the local
  ISM within 7~pc is analogous to the warm, partially
  neutral ISM that surrounds the Sun, rather than the fully ionized
  plasma that predominates within the Local Bubble.
\item[2.] With our new measurements, there are now nine M dwarf $\dot{M}$
  measurements and six meaningful $\dot{M}$ upper limits based on
  HST Lyman-$\alpha$ studies.  Of these 15 constraints, 13 are consistent
  with weak winds of $\dot{M}\leq 1$ $\dot{M}_{\odot}$.  However, even if
  generally weak, early M dwarfs seem to have
  coronal winds comparable to that of the Sun when normalized by surface
  area.  There are two M dwarfs that
  appear to have unusually strong winds:  YZ~CMi
  (M4 Ve; $\dot{M}=30 \dot{M}_{\odot}$), and GJ~15AB (M2 V+M3.5 V;
  $\dot{M}=10 \dot{M}_{\odot}$).
\item[3.] The nature of the astrospheric absorption for YZ~CMi is
  somewhat different than any observed before due mainly to the very low
  ISM flow speed of $V_{ISM}=20$ km~s$^{-1}$ seen by the star.  The
  heating in the hydrogen wall is here mainly from heat transport across
  the astropause via charge exchange rather than heating at a bow shock.
\item[4.] A new plot of $\dot{M}$ per unit surface area versus $F_X$
  for all main sequence stars, including the new M dwarf measurements,
  suggests a general increase of $\dot{M}$ with coronal activity, but
  with a roughly two order of magnitude scatter of $\dot{M}$ about
  the $\dot{M}\propto F_X^{0.77\pm 0.04}$ trend line.
  This argues that coronal activity and spectral
  type alone do not determine wind properties, with magnetic topology
  being one possible extra factor involved.  The evidence for a
  ``wind dividing line'' previously suggested for GK dwarfs at a
  $\log F_X=6.0$ is now much weaker.
\item[5.] The M dwarf wind measurements are inconsistent with the kind of
  supermassive CME-dominated wind that would be expected if the solar
  relation between flare energy and CME mass were extrapolated to
  active M dwarfs.  Thus, the flare/CME connection that seems so strong on the
  Sun \citep[e.g.,][]{ana11} does not seem to apply to M dwarfs.
  However, it is still possible that the winds that we detect for M dwarfs
  could be CME-dominated rather than quiescent in nature.
\item[6.] The new M dwarf wind constraints have important ramifications
  for the habitability of exoplanets around these stars, particularly
  the implication that CMEs may not be nearly as prevalent around such
  stars as is sometimes assumed.
\end{description}

\acknowledgments

Support for HST program GO-15326 was provided by NASA through an award from
the Space Telescope Science Institute, which is operated by the Association
of Universities for Research in Astronomy, Inc., under NASA constract
NAS 5-26555.  This research has made use of the SIMBAD database,
operated at CDS, Strasbourg, France.  This work benefited from discussions
within the international team ``The Solar and Stellar Wind Connection:
Heating processes and angular momentum loss,'' supported by the
International Space Science Institute (ISSI).  AAV acknowledges funding
from the European Research Council (ERC) under the European Union's
Horizon 2020 research and innovation programme
(grant agreement No 817540, ASTROFLOW).

\end{document}